\documentstyle[12pt,epsf]{article}
\newcommand{\la}[1]{\label{#1}}
\newcommand{\be}{\begin{equation}}
\newcommand{\ee}{\end{equation}}
\newcommand{\ba}{\begin{eqnarray}}
\newcommand{\ea}{\end{eqnarray}}

\newcommand{\nr}[1]{(\ref{#1})}
\newcommand{\msbar}{\overline{\mbox{\rm MS}}}

\newcommand{\lsi}{\raise0.3ex\hbox{$<$\kern-0.75em\raise-1.1ex\hbox{$\sim$}}}
\newcommand{\gsi}{\raise0.3ex\hbox{$>$\kern-0.75em\raise-1.1ex\hbox{$\sim$}}}
\newcommand{\lsim}{\mathop{\lsi}}
\newcommand{\gsim}{\mathop{\gsi}}
\makeatletter \@addtoreset{equation}{section} \makeatother
\renewcommand{\theequation}{\arabic{section}.\arabic{equation}}
\begin{document}

\setlength{\baselineskip}{0.6cm}
\newcommand{\nn}{\nonumber}
\newcommand{\tr}{{\rm Tr\,}}
\newcommand{\fr}[2]{{\frac{#1}{#2}}}
\newcommand{\bmu}{\bar{\mu}}
\newcommand{\Lf}{L_f(\bar{\mu})}
\newcommand{\Lb}{L_b(\bar{\mu})}
\newcommand{\tb}{\tan\!\beta}
\newcommand{\pf}{\frac{1}{16\pi^2}}
\newcommand{\hh}{H^{\dagger}H}
\newcommand{\uu}{U^{\dagger}U}
\newcommand{\hd}{H^{\dagger}}
\newcommand{\thd}{\tilde{H}^{\dagger}}
\newcommand{\sbb}{\sin\!2\beta\,}
\newcommand{\cbb}{\cos\!2\beta\,}
\newcommand{\ssb}{\sin^2\!\beta\,}
\newcommand{\ccb}{\cos^2\!\beta\,}
\newcommand{\mtR}{m_{\tilde{t}_R}}
\newcommand{\mtRc}{m_{\tilde{t}_R,c}}
\newcommand{\figysize}{16.0cm}
\newcommand{\figtopspace}{\vspace*{-1.5cm}}
\newcommand{\figbottomspace}{\vspace*{-5.0cm}}

\begin{titlepage}
\begin{flushright}
HD-THEP-96-56\\
hep-ph/9612364\\
December 14, 1996
\end{flushright}
\begin{centering}
\vfill

{\bf 
The 2-loop MSSM finite temperature effective potential
with stop condensation}
\vspace{1cm}

D. B\"odeker\footnote{cq1@ix.urz.uni-heidelberg.de}, 
P. John\footnote{p.john@thphys.uni-heidelberg.de}, 
M. Laine\footnote{m.laine@thphys.uni-heidelberg.de} and
M.G. Schmidt\footnote{m.g.schmidt@thphys.uni-heidelberg.de} \\

\vspace{1cm} {\em 
Institut f\"ur Theoretische Physik, 
Philosophenweg 16, 
D-69120 Heidelberg, Germany}

\vspace{2cm}

{\bf Abstract}

\vspace{0.5cm}
We calculate the finite temperature 2-loop effective potential in the
MSSM with stop condensation, using a 3-dimensional effective theory.
We find that in a part of the parameter space, a two-stage electroweak
phase transition appears possible. The first stage would be the
formation of a stop condensate, and the second stage is the transition
to the standard electroweak minimum. The two-stage transition could
significantly relax the baryon erasure bounds, but the parameter space
allowing it ($m_H \lsim 100$ GeV, $\mtR \sim 155-160$ GeV) is not very
large. We estimate the reliability of our results using
renormalization scale and gauge dependence. Finally we discuss some
real-time aspects relevant for the viability of the two-stage scenario.
\end{centering}

\vspace{0.3cm}\noindent

\vfill \vfill
\noindent

\end{titlepage}


\section{Introduction}
\la{intro}

The electroweak (EW) phase transition in the early universe has been
under intense investigation in recent years (for reviews, 
see~\cite{rs,rsbb}). 
Indeed, the Standard Model (SM) offers the three necessary ingredients
of Sakharov for baryogenesis~\cite{rs,krs}. However, careful studies 
have shown that in the SM (i) it may be somewhat difficult 
to produce enough baryons during the EW phase transition, 
due to the limited amount of CP-violation available and (ii) even
if a $B+L$  asymmetry is produced, it is washed out afterwards
by the sphaleron effects. This is because the
phase transition is only very weakly of first order for the
experimentally allowed Higgs masses $m_H\gsim 60$ GeV, 
and even vanishes for $m_H\gsim m_W$~\cite{klrs2}.
 
Thus one has to return to speculative models of baryogenesis
in the very early stages of the universe marked by grand unification
and related topics --- or one has to consider variants of the SM.
The SM has been and still is incredibly successful in
explaining experiments though its limited range is generally accepted.
Supersymmetric extensions are today the only way to
maintain perturbative calculability at higher scales
and the practical success of the SM. 
Furthermore, from the theoretical point of view supersymmetry (SUSY)
appears to be the only known attractive way of including 
quantum gravity through supersymmetric string theory. 
Still, SUSY has to be detected in experiments!
 
The most simple SUSY model is the minimal supersymmetric extension (MSSM) of
the SM: here the only new particles proposed are the superpartners
with known (SM) couplings and an additional Higgs doublet.
This is a very strong restriction in view of the multitude of possibilities
offered, e.g., by string theory, but even in this simple case the
necessary soft SUSY breaking terms introduce a huge number
of free parameters. Fortunately we are interested in the behaviour
of such a model at the high temperatures of a (modified)
EW phase transition, in which case the problem 
simplifies. The new fermionic superpartners,
the gauginos and higgsinos, become even more massive in the heat
bath and can be integrated out. They may
be important, however, for introducing strong CP-violating
effects. The scalar quark partners, if heavy enough, can also
be integrated out. Then the remaining nonzero Matsubara modes
and heavy zero modes 
contribute to a 3-dimensional (3d) effective 
action like in the standard case~\cite{g}. 
``Integrating out'' in the above means
matching parameters in the comparison of a 4d and 3d
set of amplitudes as in~\cite{fkrs1,klrs1,bn}. The effective theory
for the light modes contains Higgs and gauge fields. If  
some combination of the two Higgs doublets is heavy compared
with the scale $g^2T$, it can also be integrated out, 
resulting in an effective action of the
same general form as for the SM~\cite{klrs1,ck,lo,ml}. 
This case leads to conclusions which are not much
more optimistic than in the SM~[9--14]: 
for generic parameters,   
the phase transition would be at most relatively weakly 
of first order for Higgs masses $m_H \gsim 60-70$ GeV.

However, recently it has been argued that a rather 
light ($\mtR\lsim m_t$) stop, the superpartner of the right-handed top, 
and low values of $\tan\beta$,
allow for a strongly first-order phase transition even 
at Higgs masses $m_H\sim m_W$~\cite{cqw,dggw}. 
Furthermore, 
the influence of 2-loop contributions to the effective
potential was observed to be large due to the strong interactions
of the stop entering at the 2-loop level, 
strengthening the transition~\cite{e}. 
  
In this paper we investigate the MSSM with a light stop
in some detail. In particular, 
we study the possibility of a two stage transition
in which  at some temperature the stop acquires a nonzero expectation value
while the electroweak symmetry is still unbroken. In the second
stage the universe would make a transition from the charge and colour
breaking (CCB) minimum to the standard one. In Ref.~\cite{cqw}
it was argued that a CCB minimum does not develop during the 
electroweak phase transition as long as  $\mtR$ satisfies the 
bound necessary for vacuum stability at zero temperature;
here we demonstrate that stop condensation may be possible
in some part of the parameter space after all, leading
to a two-stage transition in which the 
standard EW minimum is the global one at low temperatures. 
Such a transition would relax the baryon erasure bounds
and allow for large Higgs masses, $m_H \lsim 100$ GeV.

We compute the effective potential for the Higgs and stop fields
at two loops in 3d, working in an arbitrary 
background gauge. The gauge and renormalization scale
dependence of our results should give an estimate of the size
of higher loop contributions. 
Finally, assuming that a two-stage transition takes place, 
we discuss some issues concerning the real-time history of
such a transition. Previously, two-stage transitions 
(in two Higgs doublet models) have been considered in~\cite{ts}.

The plan of the paper is the following. 
In Sec.~\ref{3dtheory} we discuss the 3d effective
theory describing the EW phase transition in the MSSM.
Sec.~\ref{2loop} contains the derivation of the 2-loop
effective potential in the 3d theory. Implications 
from the 2-loop potential for
the phase diagram of MSSM 
are presented in Sec.~\ref{numerical}. In Sec.~\ref{nucleation}
we address some real-time aspects of the finite temperature
phase transitions found, and the conclusions 
are in Sec.~\ref{conclusions}. Some details
related to Secs.~\ref{3dtheory}, \ref{2loop}
are in the two appendices.

\section{The effective 3d theory and its parameters}
\la{3dtheory}

The effective 3d theory relevant for the EW
phase transition in MSSM is~\cite{ml}
\ba
L & = &
\fr14 F^a_{ij}F^a_{ij}+\fr14 G^A_{ij}G^A_{ij} 
+(D_i^w H)^\dagger(D_i^w H)+m_{H3}^2 H^\dagger H+
\lambda_{H3} (H^\dagger H)^2 \nn \\
& + & (D_i^s U)^\dagger(D_i^s U)+m_{U3}^2U^\dagger U+
\lambda_{U3} (U^\dagger U)^2
+ \gamma_3 H^\dagger H U^\dagger U. \la{Uthe}
\ea
Here $D_i^w=\partial_i-i g_{W3}\tau^a A_i^a/2$ and
$D_i^s=\partial_i-i g_{S3}\lambda^A C_i^A/2$ are the 
SU(2) and SU(3) covariant derivatives, 
$g_{W3}$ and $g_{S3}$ are the corresponding 3d gauge couplings, 
$H$ is the Higgs doublet and $U$ is the right-handed stop field. 
This theory is an effective theory for the thermodynamics of the 
EW phase transition in MSSM, provided that the parameters 
are suitably fixed in terms of the temperature $T$ and the 
zero-temperature physical parameters of the theory~\cite{lo,ml} 
(see also below). The theory in~\nr{Uthe}
should be particularly
useful for the case that the right-handed stops are
light, $m_{\tilde{t}_R} < m_t$,
which has recently attracted a lot of attention 
since the EW phase transition appears then to be 
strong enough for baryogenesis~\cite{cqw,dggw,ck,ml}. 
For a heavier stop, $\mtR > m_t$, the $U$ field can 
be integrated out and the effective theory is just 
the 3d SU(2)$+$Higgs model as in the Standard Model~\cite{ck,lo,ml}.

The reasons why it may be convenient (even in perturbation theory)
to study the theory~\nr{Uthe} rather than to directly
calculate the effective potential in 4d, are: 

1. The use of~\nr{Uthe} factorizes the problem into two parts.
The first part, the derivation of~\nr{Uthe}, is purely perturbative
and free of IR-problems. The second part is the analysis of~\nr{Uthe}
in 3d and is subject to the usual IR-problems at finite $T$.
Thus the IR-problems can be studied in a simplified setting.

2. The effective theory~\nr{Uthe} has a certain degree of
universality, since the same 3d theory (with different
parameter values) 
arises for many different 4d theories. 
Thus the IR-problems of many 4d theories can be 
studied once and for all using~\nr{Uthe}.

3. The construction of~\nr{Uthe} automatically implements the 
daisy resummation procedure required for the mass parameters
at finite temperature, as well as resummations for all the 
couplings appearing in the theory (establishing the scale 
at which they are to be evaluated). In addition, there
are resummations related to heavy modes in 3d. 
Apart from the daisy resummation, these
resummations are usually not implemented in 
direct calculations in 4d. Yet their effect may be
numerically non-negligible: e.g., the finite temperature 
strong gauge coupling is much smaller than $g_S^2(m_Z) T$.

4. One may eventually want to perform lattice simulations 
of the EW phase transition in the MSSM by using~\nr{Uthe}. Note 
that while the simulations in the Standard Model 
indicated that perturbation theory is surprisingly 
reliable for strong transitions~\cite{klrs2,leip,desylattice}, 
this may not necessarily hold in MSSM with two light
scalar fields. In fact, we find that  
the renormalization scale and gauge
dependence of the results is much larger
than in the SM indicating that the convergence
of the 2-loop perturbative expansion is worse. 
It should also be noted that there are always
massless SU(3) gauge excitations related to 
an unbroken SU(2) subgroup in perturbation theory 
(even if the $U$-field has an expectation value), 
unlike in the SM. 
If lattice simulations are ever made, 
then it is useful to have perturbative results in the 
same theory for comparison.

These benefits are slightly shadowed by the fact that at 
present there is no accurate derivation of the parameters
in~\nr{Uthe} available for a wide range of 4d zero temperature
parameters. While this derivation is 
straightforward and parallels the ones in~\cite{ck,lo,ml}, 
it is complicated by the fact that the whole
mass spectrum of the MSSM enters 
through relatively large radiative corrections, 
especially in the strongly interacting sector. Moreover 
(as is discussed in more detail in \ref{app1}), 
the derivation should be made at 2-loop level as concerns
the mass parameters, if one is studying a two-stage
scenario where an accurate comparison of two critical temperatures
is needed. This is because it is only a 2-loop calculation
that fixes the scale appearing in the thermal screening
terms ($\sim g_S^2T^2$).
While a 2-loop derivation is 
quite doable~\cite{klrs1}, we will not go into it here.
Instead we make a 1-loop derivation for a particular
region of the parameter space, and an 
order of magnitude estimate of
the 2-loop scale factors appearing. This derivation 
should illustrate with reasonable accuracy the real situation, 
and allows us to address some interesting problems: 

1. In~\cite{cqw}, it was found that a two-stage transition
where the right-handed stop field ($U$)
gets an expectation value should not be 
allowed, since then the CCB minimum would be the global
one and one would remain there until the present day 
(a two-stage transition was 
argued to be ruled out in~\cite{ccb}, as well, 
since it was found that there is no second order 
transition by which to get back to the standard EW minimum). 
Here we try to be somewhat
more accurate in a particular region of the parameter space, 
and we demonstrate that a two-stage
transition might be possible after all. While there
remain uncertainties in our calculation and thus it is not clear
whether this case is eventually realized, it is an interesting 
prospect and hence in our opinion worth studying. 

2. In~\cite{e} it was found that for small right-handed stop
masses, the 2-loop corrections in the effective potential 
for the $H$-direction are large, making the transition 
stronger. 
Here we calculate the 2-loop potential in a general
$U, H$-background. We confirm the 2-loop effect found in~\cite{e},
find similar large 2-loop corrections in 
the $U$-direction as well, and discuss the reliability
of the statements based on these corrections. 
It should be stressed that the large 2-loop corrections
are IR-sensitive 3d effects and thus independent of
the 4d parameters used in the derivation of the 
effective theory. 

Since the derivation of the parameters in~\nr{Uthe} is 
a problem which completely factorizes from the calculation
of the 2-loop effective potential in 3d, we will discuss
this derivation separately in \ref{app1}. 
Let us here just summarize the 
zero temperature vacuum parameters used:
the CP-odd Higgs mass $m_A$ is assumed for simplicity 
to be large ($m_A \gsim 300$ GeV), the squark mixing parameters 
are very small, the running top mass is $m_t\sim 170$ GeV and
the left-handed squark mass parameter is relatively small, 
$m_Q\sim 300$ GeV. We then present the results as a 
function of $\tb$ (or equivalently the Higgs mass $m_H$) 
and the right-handed stop mass $\mtR$.

\section{The 2-loop potential and physical observables}
\la{2loop}

The effective theory~\nr{Uthe} contains two fundamental
representation Higgs fields: the Higgs doublet $H$ interacting
via the electroweak SU(2) group, and the right-handed stop 
colour triplet
$U$ interacting via the strong SU(3) group. We have calculated
the 2-loop effective potential of this theory in a general
$H, U$ background. After gauge fixing, there remains
a global symmetry in the theory, so the effective potential
only depends on 
\be
H^\dagger H \equiv \frac{\phi^2}{2},\quad
U^\dagger U \equiv \frac{\chi^2}{2}.
\ee
In the 3d theory, the dimension of $\phi, \chi$ is GeV$^{1/2}$
after a trivial rescaling with $T$; we nevertheless often
use for clarity the 4d dimensions, so that  
\be
\frac{\phi_{\rm 4d}}{T}=\frac{\phi_{\rm 3d}}{\sqrt{T}},
\ee 
and similarly for $\chi$.

It should be noted that within the 3d theory~\nr{Uthe},
the temperature $T$ does not appear explicitly. It is hidden 
in the parameters of the 3d theory, for instance
in $g_{S3}^2 = g_S^2 T$. All the dimensionful 3d observables could 
hence be expressed in terms of $g_{S3}^2$. However, to keep
the connection to 4d physics clear and since the values
of $g_{S3}^2, T$ are numerically close to each other, we 
will rather express the results in terms of $T$.

We have calculated the effective potential $V(\phi,\chi)$ 
in a background field gauge with two gauge parameters, 
$\xi$ related to SU(2) and $\zeta$ related to SU(3). 
The applicability of the background field gauge to 
the present context has been discussed in~\cite{kls1}.
We display our results  
in the Landau gauge $\xi,\zeta = 0$, and
other values of the gauge parameters are used in estimating
the convergence of the calculation (see below). 
For generality, we have replaced SU(3) by SU($N$)
in the calculation and we also keep a general dimension $d$
in the formulas before the evaluation of the final 
two integrals~\nr{intI}, \nr{intH}, to which all others reduce.
We work throughout in the $\msbar$ scheme with
the scale parameter $\bmu$. The details of the calculation are 
in \ref{app2}.

{}From the effective potential we extract several gauge fixing 
independent
physical quantities. The critical temperature $T_c^\phi$ for the
transition from the origin to the standard electroweak (EW) minimum 
$\phi_{\rm min}$ is
defined such that at $T_c^\phi$, 
\be 
V(0,0)=V(\phi_{\rm min},0).  
\ee
Similarly one finds the critical temperature $T_c^\chi$ for the
transition from the origin to the charge and colour breaking
(CCB)
minimum $\chi_{\rm min}$ in the $U$ direction, and $T_c^{\chi\to\phi}$
for the transition from the CCB minimum to the EW minimum. 

The strength of the phase transition with respect to the 
sphaleron rate is characterized by the vacuum expectation 
value $\phi_{\rm min}/T$ in the broken phase. This is a
gauge dependent quantity; we calculate it in the 
Landau gauge. A gauge independent characterization of 
the discontinuity could be obtained with the 
expectation values of the composite
operators $H^\dagger H, U^\dagger U$~\cite{pdp}:
\ba
\Delta\langle H^\dagger H\rangle & = &  
\frac{d}{dm_{H3}^2}\biggl[
V(\phi_{\rm min},0)-V(0,0)
\biggr],\nonumber \\
\Delta\langle U^\dagger U\rangle & = &  
\frac{d}{dm_{U3}^2}\biggl[
V(0,\chi_{\rm min})-V(0,0)
\biggr]. \la{hdh}
\ea
However, numerically $\Delta\langle H^\dagger H\rangle\sim
\phi_{\rm min}^2/2$, $\Delta\langle U^\dagger U\rangle\sim
\chi_{\rm min}^2/2$, so that we do not separately discuss
the quantities in~\nr{hdh}. 

The latent heat is defined by
\be 
L=T_c\frac{d}{dT}\left.\biggl(p_{h}(T)-p_{l}(T)\biggr) 
\right|_{T=T_c}, \la{latent}
\ee 
where $p_{h}(T)$ ($p_{l}(T)$) 
is the pressure $p(T)=-V({\rm min})$ in the high (low)
temperature phase (note that we mostly write $V(\phi,\chi)$
in 4d units; the relation to 3d is 
$V_{\rm 4d}=TV_{\rm 3d}$,
apart from field independent terms). 
For the surface tension we use the leading order
expression 
\be \sigma = \int_{\varphi_1}^{\varphi_2}\!
d\varphi\,\sqrt{2 [V(\varphi)-V(\varphi_1)]} ,
\label{sigmaint}
\ee 
where $\varphi=(\phi,\chi)$,
$d\varphi$ is the length element in the field space, 
and the path is the one that minimizes the result
(this can be derived by extremizing the bounce action at~$T_c$). 
Perturbative corrections to the surface tension 
from the derivative terms in the SU(2)$+$Higgs model
were discussed in~\cite{desy,kls2}.

It is important to estimate the reliability of the 
perturbative results obtained. To this end, one can 
use the fact that the physical observables derived
are gauge- and $\bmu$-independent to the order to which 
they have been calculated. The remaining dependence
then gives an estimate of the magnitude of higher-order
corrections. For the gauge dependence, we vary $\xi,\zeta$
from zero to $\sim 1$. For the $\bmu$-dependence, we compare
results from the choice $\bmu=T$ with results
from a renormalization group (RG) improved 
optimized choice~\cite{fkrs1}. The optimization condition is that
the 2-loop contribution to $V'(\varphi)$ vanishes, 
and it gives a field dependent scale parameter $\bmu(\varphi)$
which should be proportional to some (non-linear) combination of 
the mass scales contributing to $V(\varphi)$. The simple 
choice $\bmu=T$ is also supposed to 
reflect the mass scales of the problem, since 
the non-perturbative 3d masses should be of 
order $g_{W3}^2, g_{S3}^2$~\cite{klrs2,leip,knpr,phtw,dkls}.

\section{Numerical results}
\la{numerical}

The 2-loop effective potential $V(\phi,\chi)$
for a particular choice of the parameters 
($\tb=5$, $\mtR=158.3$ GeV, $\bmu=T$)
at $T=T_c=92.43$ GeV
is shown in Fig.~\ref{3dfig}. This choice
corresponds to a special case 
in which there is a simultaneous first order
transition in both directions. In general, 
the transitions take place at different 
temperatures $T_c^\phi, T_c^\chi$, and
there is a third temperature $T_c^{\chi\to\phi}$
at which the broken minima are at equal heights. 

\begin{figure}[tb]

\vspace*{-0.5cm}

\epsfysize=9.0cm
\centerline{\epsffile{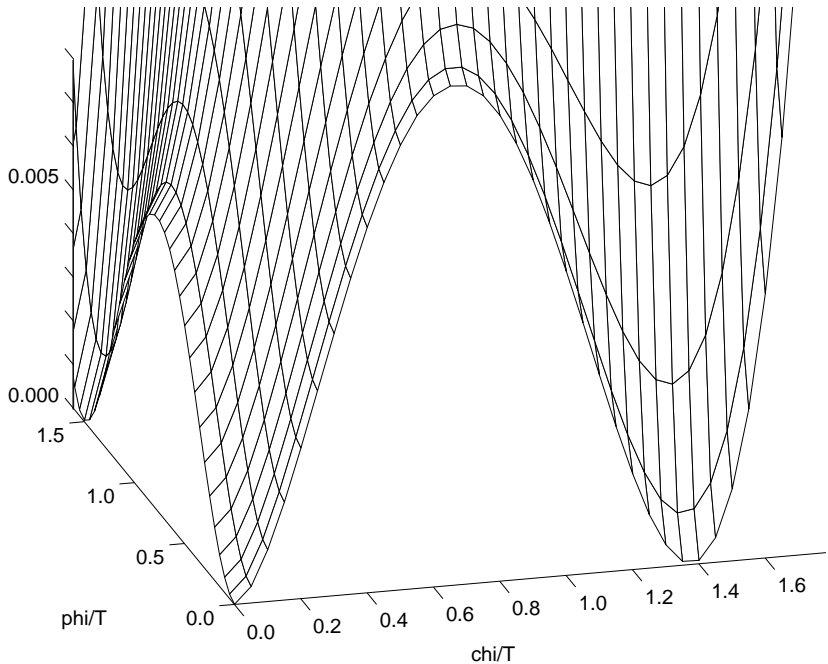}}
\caption[a]{The 2-loop effective potential 
at $T=T_c^\phi=T_c^\chi=T_c^{\chi\to\phi}$ 
for $\tb=5$ ($m_H\sim 92$ GeV), $\mtR=158.3$ GeV, $\bmu=T$, 
$\xi=\zeta=0$.}
\la{3dfig}
\end{figure}

The 2-loop potential corresponding to the parameters of Fig.~\ref{3dfig}
is displayed more precisely in the 
$\phi$- and $\chi$-directions in Fig.~\ref{veff}, together
with the 1-loop potentials at the corresponding critical
temperatures. One can observe several things. First, looking 
at the 1- and 2-loop potentials at the same $\bmu=T$, one sees
that the 2-loop effects are large and seem to make the transition
much stronger. For the $\phi$-direction this was observed
in~\cite{e}, and the reason for the strengthening was tracked
down to 2-loop graphs involving the strongly interacting stops.  
{}From Fig.~\ref{veff} one sees that there are similar large
2-loop corrections in the $\chi$-direction, as well.

\begin{figure}[p]

\vspace*{-1.5cm}
 
\centerline{\hspace{-3.3mm}
\epsfxsize=11cm\epsfbox{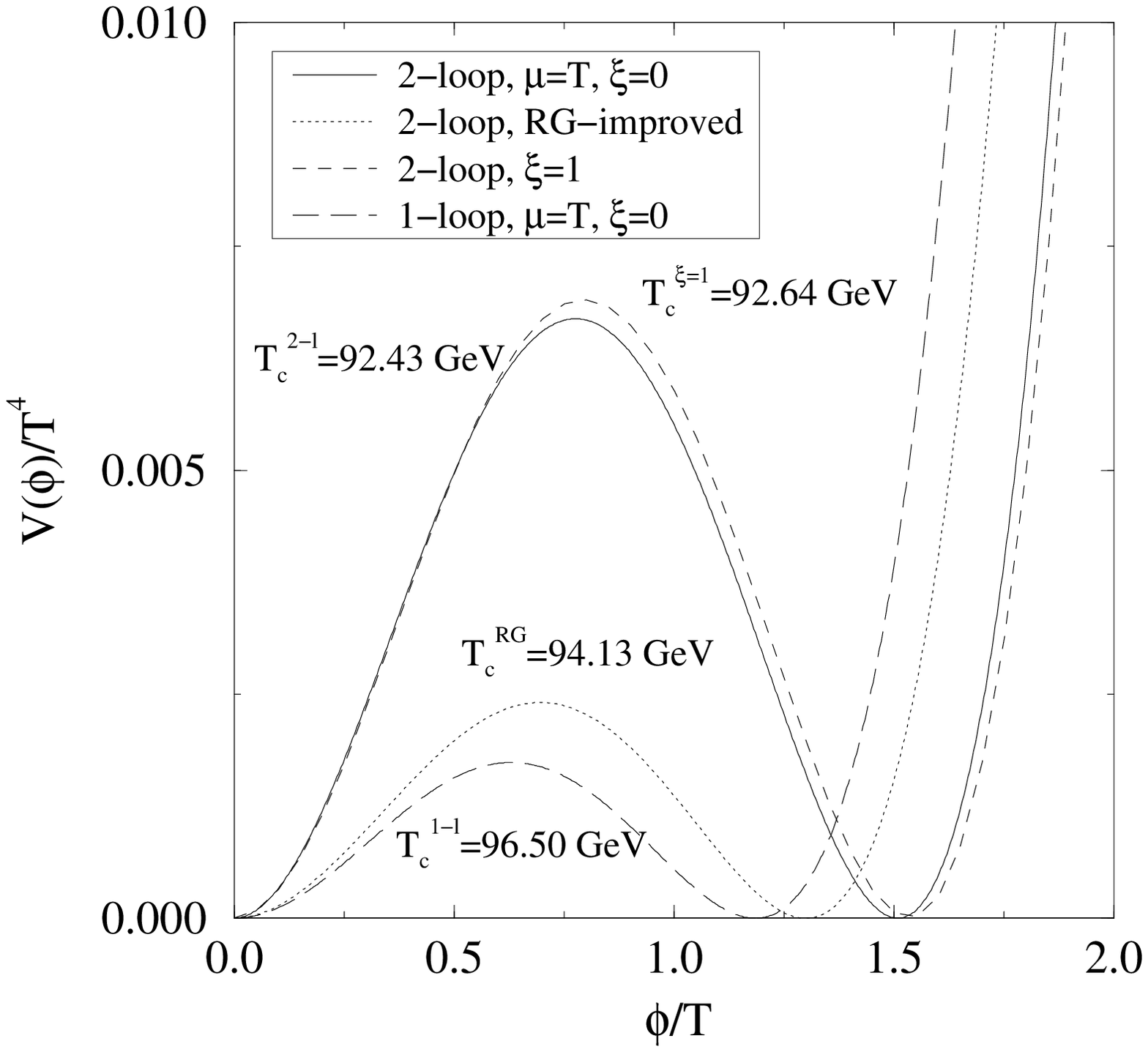}}

\vspace*{-6cm}

\centerline{\hspace{-3.3mm}
\epsfxsize=11cm\epsfbox{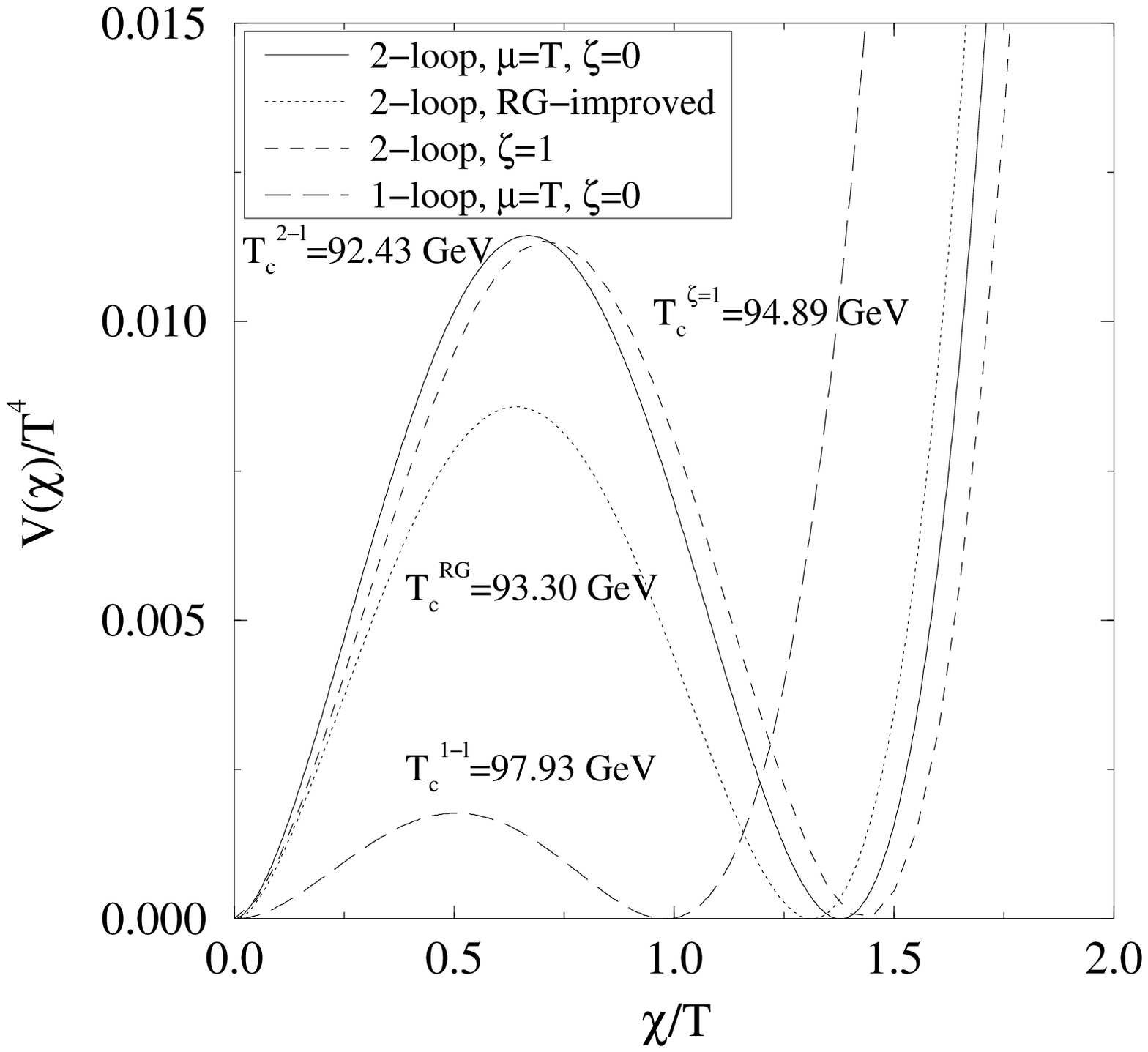}}
 
\vspace*{-5cm}
 
\caption[a]{The 1-loop and 2-loop effective potentials
at the corresponding critical temperatures, 
for $\tb=5, \mtR=158.3$ GeV. It is seen
that there may be a relatively large uncertainty
in the prediction for the surface tension due to the
large $\bmu$-dependence, especially for the $\phi$-direction. 
The figure has been drawn at the special point $\mtR=\mtRc$
so that the critical temperatures $T_c^{\rm 2-l}$ of the two transitions
are equal (see Fig.~\ref{Tc}). The $\bmu$-dependence
is especially large at this point, since the mass parameters 
$m_{H3}^2(\bmu)$,
$m_{U3}^2(\bmu)$ are close to zero so that the relative change
can be significant.}
\la{veff}
\end{figure}

Second, comparing the 2-loop potentials at different
choices of $\bmu$ ($\bmu=T$ vs.\ RG-improved $\bmu$), 
one can see that the reliability of the 
2-loop results is questionable, especially for the 
$\phi$-direction. The large $\bmu$-dependence arises
from the $\bmu$-dependence of $m_{U3}^2(\bmu)$ (see Eq.~\nr{mmU3mu})
in the 1-loop corrections, and would be cancelled by 
3-loop graphs involving strong interactions. 
Thus the 3-loop contributions are quite important. 
The large $\bmu$-dependence seen here is in contrast
to what has been observed in the pure SU(2)+Higgs model~\cite{fkrs1}.
At the same time, even in the SU(2)+Higgs model
non-perturbative results for the surface tension 
(which is characterized by the size of the bump 
between the minima, see~\nr{sigmaint})
were significantly smaller than 2-loop perturbative results 
for weaker transitions~\cite{klrs2,kls1}. Thus 
one can say that for the surface tension 
perturbative results are expected to 
give only an order of magnitude
estimate. For the critical temperature the $\bmu$-dependence
in Fig.~\ref{veff} is somewhat
smaller, so the existence of the two transitions and 
their crossing point is on a firmer basis. 
For the vevs and latent heats the $\bmu$-dependence is 
also relatively
smaller than for the surface tension, so the 
perturbative estimates should produce the correct
qualitative features.   

The gauge dependence of $V(\phi,\chi)$ (obtained by varying
$\xi,\zeta$ from zero to unity) is also 
shown in Fig.~\ref{veff}.
It differs from the $\bmu$-dependence in being 
much larger in the $\chi$-direction, especially
for $T_c$. The reason for the difference is that 
the $\bmu$-dependence arises from scalar 
(in particular stop) excitations 
affecting also the $\phi$-direction, 
whereas the gauge dependence arises from vector excitations. 
For quantities other than $T_c$ the gauge dependence
appears to be smaller.  It should be noted 
that the gauge dependence (as well as the $\bmu$-dependence)
is significantly smaller at the 2-loop than at the 1-loop level. 
The error bars containing an estimate of both 
the 2-loop gauge and 
$\bmu$-dependence for 
the critical temperatures are shown in Fig.~\ref{Tc}.

\begin{figure}[tb]

\figtopspace

\epsfysize=\figysize
\centerline{\epsffile{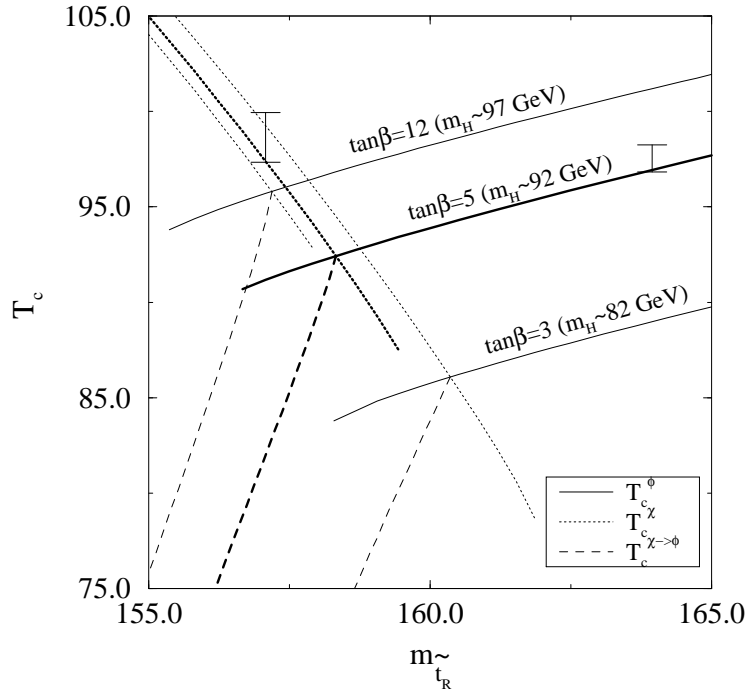}}

\figbottomspace

\caption[a]{
The critical temperatures of the three transitions 
as a function of $\mtR$ for $\tb=3,12$ (thin lines) 
and $\tb=5$ (thick lines).
The two-stage transition would
take place to the left of the crossing point of 
the three critical curves; it is seen that a two-stage
transition is possible, but there is not very
much parameter space for it. The continuations of $T^\phi_c$
to the left of the crossing point, and $T^\chi_c$ to the right of it, 
roughly represent the region of metastability of the transition 
with a higher $T_c$.
Error bars indicate the 
dependences on $\bmu, \xi, \zeta$ (for $\tb=5$),
as explained in the text; the actual curves
correspond to the Landau gauge and $\bmu=T$.
In terms of the parameter 
$m_U^2=\mtR^2-m_t^2$, the $x$-axis is 
from $-(70$ GeV)$^2$ to $-(40$ GeV)$^2$.}
\la{Tc}
\end{figure}

The dependence of the critical temperatures on
$\mtR$ and $\tb$ is also shown in Fig.~\ref{Tc}.
The general pattern is the following: For a given $\tb$, 
there is a certain critical value of $\mtR$, for instance
$\mtRc=158.3$ GeV for $\tb=5$. If $\mtR > \mtRc$, 
the critical temperature $T_c^\phi$ of the normal ($0\to\phi$)
EW phase transition is higher than the critical temperature $T_c^\chi$
of the CCB transition ($0\to\chi$). In this case
the EW phase transition proceeds in the normal
way when the universe has cooled down to $T_c^\phi$. 
If $\mtR < \mtRc$, on the contrary, then 
$T_c^\chi > T_c^\phi$ and the universe ends up in 
the CCB minimum (provided that the first order
phase transition is weak enough compared with 
the expansion rate of the universe). However, for the parameter
values we are using, we find that the CCB minimum is not the global
one at much lower temperatures. Instead, there is another
first order phase transition at $T_c^{\chi\to\phi}<T_c^{\chi}$
in which the normal EW minimum becomes the global one. 
Thus, if this latter transition is weak enough to 
take place within the cosmological time scales available, 
one ends up finally in the normal EW minimum. The 
vacuum expectation value $\phi_{\rm min}$ can 
easily be large $\phi_{\rm min}/T_c \gg 1$ 
after the latter transition (Fig.~\ref{vev}),
so that the sphaleron rate 
would be very effectively switched off.

\begin{figure}[tb]

\figtopspace

\epsfysize=\figysize
\centerline{\epsffile{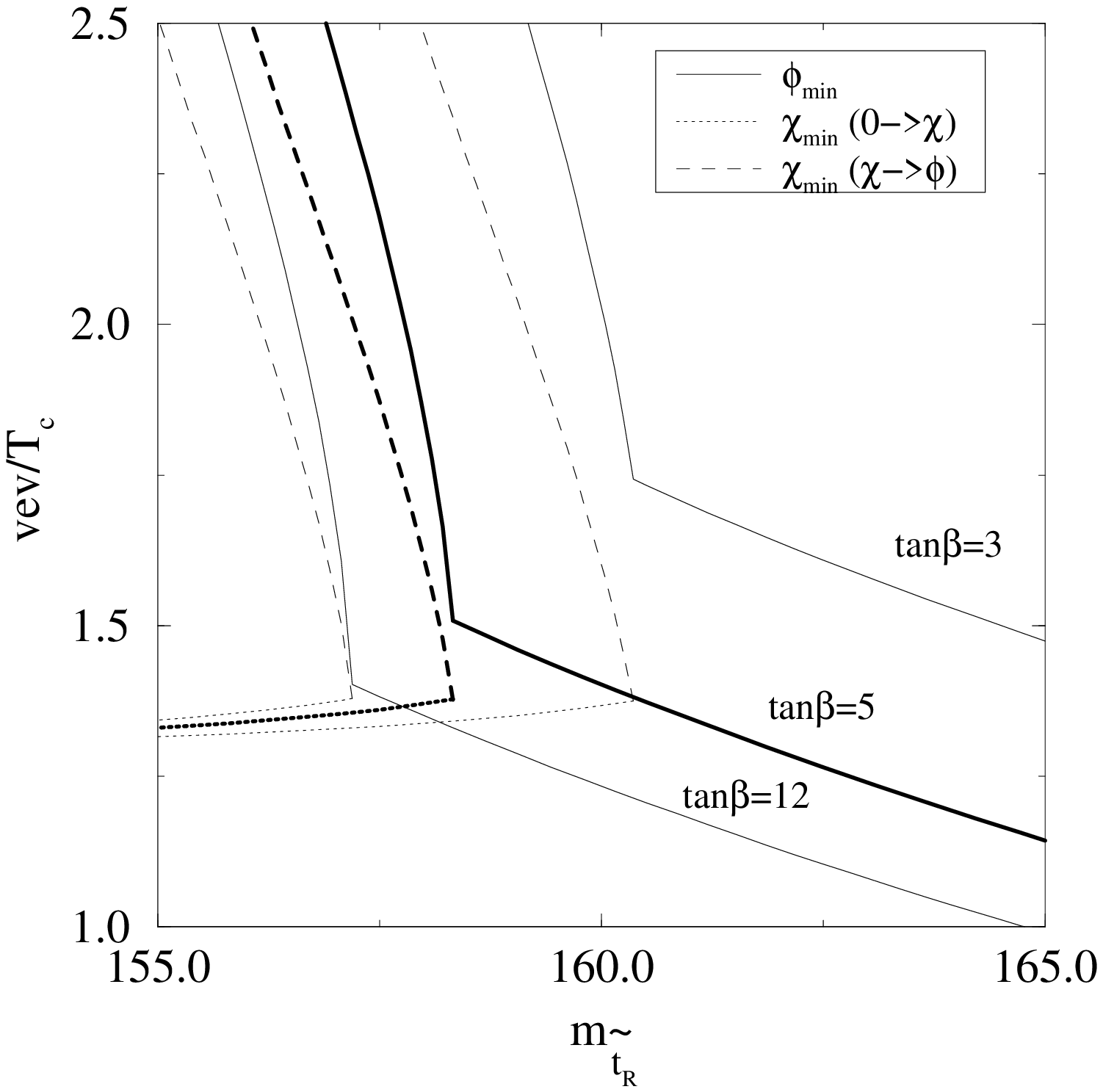}}

\figbottomspace

\caption[a]{
The expectation values 
$\phi_{\rm min}/T_c$, 
$\chi_{\rm min}/T_c$ 
of the three transitions (see Fig.~\ref{Tc})
in the Landau gauge 
as a function of $\mtR$ for $\tb=3,5,12$. The broken 
phase transition from $\chi_{\rm min}>0$ to 
$\phi_{\rm min}>0$ is seen to easily lead to 
very large values of $\phi_{\rm min}$. For clarity
the metastable branches are not shown here.}
\la{vev}
\end{figure}

In Fig.~\ref{Tc} we have also shown the metastable
branches of the EW ($0\to\phi$) and CCB ($0\to\chi$) transitions. 
These could be relevant if the transition with the higher $T_c$
is so strong that it has not taken place before the temperature
has cooled down to the $T_c$ of the lower 
transition (see Sec.~\ref{nucleation}). 
When one goes far enough into the metastability region, then 
the higher transition definitely does take place since it would
have reached the barrier temperature by the time of the lower $T_c$.

\section{Real-time history}
\la{nucleation}

In the previous Section, we demonstrated that the phase
diagram of the EW matter described by the MSSM may be such
that the cosmological EW phase transition could take place
in two stages. However, the transitions occurring would
be of first order, so one has to discuss the amount
of supercooling taking place and the possible reheating due to 
the latent heat released, as well, to see how the transitions
would really proceed in the early universe. We will here
make some simple estimates on the presumable real-time events.

\begin{figure}[tb]

\figtopspace

\epsfysize=\figysize
\centerline{\epsffile{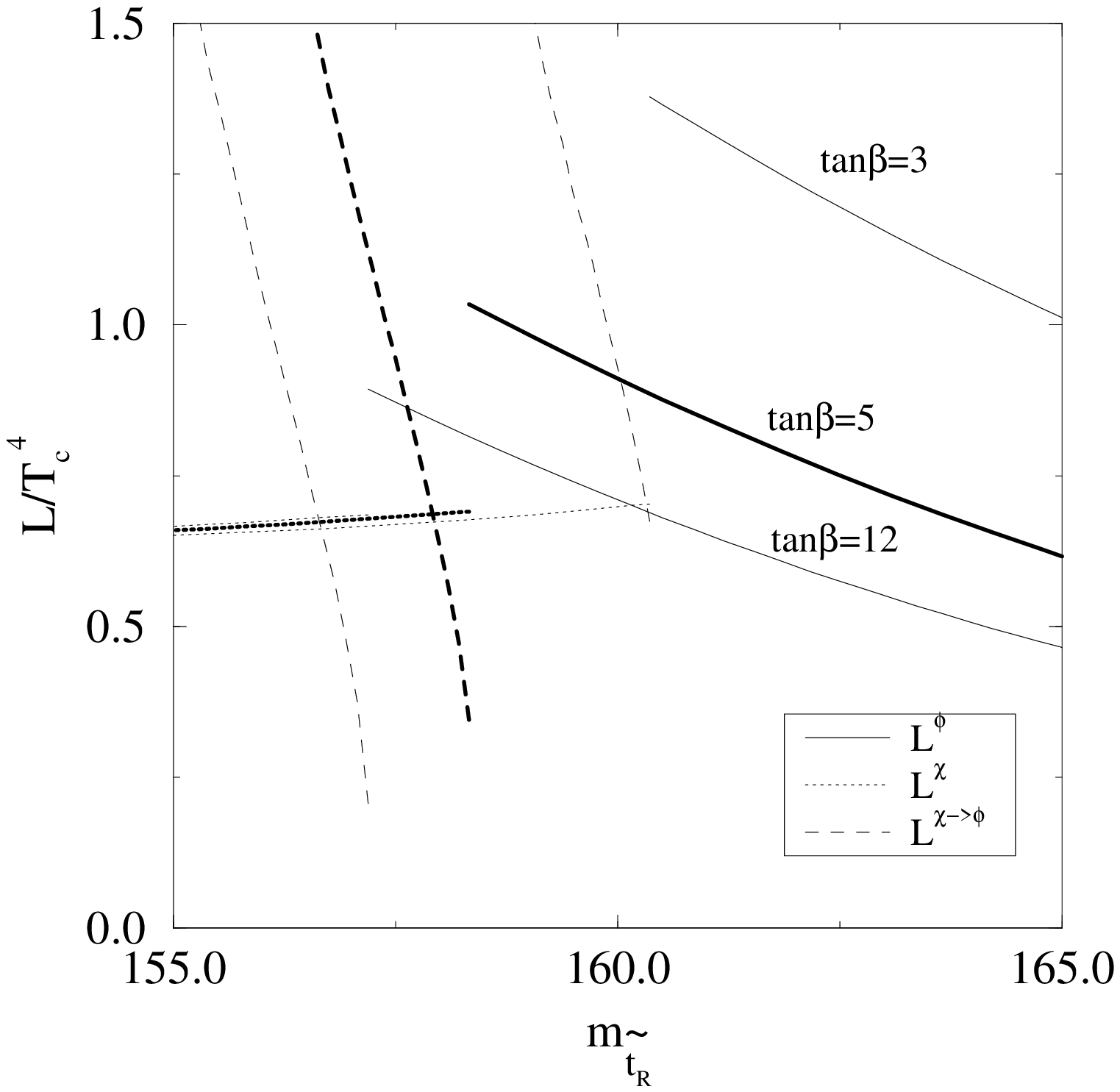}}

\figbottomspace

\caption[a]{
The latent heats of the transitions 
as a function of $\mtR$ for $\tb=3,5,12$. Note that at $\mtRc$,
$L^{\chi\to\phi}=L^{\phi}-L^{\chi}$. For smaller $\mtR$, 
$L^{\chi\to\phi}$ grows rapidly.
Metastable branches are not shown.}
\la{L}
\end{figure}

The parameters needed for the real-time 
estimates are the latent heat $L$ 
and the surface tension $\sigma$, scaled with
powers of $T_c$: $\hat{L}\equiv L/T_c^4$, 
$\hat{\sigma}\equiv \sigma/T_c^3$. 
Indeed, the small supercooling formula~\cite{kk,eikr} for the 
nucleation temperature $\hat{T}_n=T_n/T_c$ is 
\be
1-\hat{T}_n\sim 0.4 
\frac{\hat{\sigma}^{3/2}}{\hat{L}}, \la{nucl} 
\ee
assuming that in the high temperature phase $p(T) \propto T^4$.
This formula can be expected to be valid only when 
$1-\hat{T}_n \ll 1$, 
and it usually breaks down by 
underestimating the value of $1-\hat{T}_n$.
Another quantity of interest is the amount 
of heating that takes place after nucleation
as the latent heat is released. Since the bubbles
tend to fill the universe in a time scale very small 
compared with the time scale of expansion~\cite{desy,eikr}, 
the reheating temperature $T_r$ can be estimated from
\be
e_h(T_n)=e_l(T_r), \la{reheat}
\ee
where $e_h(T), e_l(T)$ are the energy densities
in the high and low temperature phases, respectively. 
{}From~\nr{reheat} one finds that reheating
to the critical temperature ($T_r>T_c$) takes place
roughly if $\hat{L}\gsim 8\hat{\sigma}^{3/4}$, assuming 
that the effective number of massless degrees of freedom 
is $\sim 110$.

\begin{figure}[tb]

\figtopspace

\epsfysize=\figysize
\centerline{\epsffile{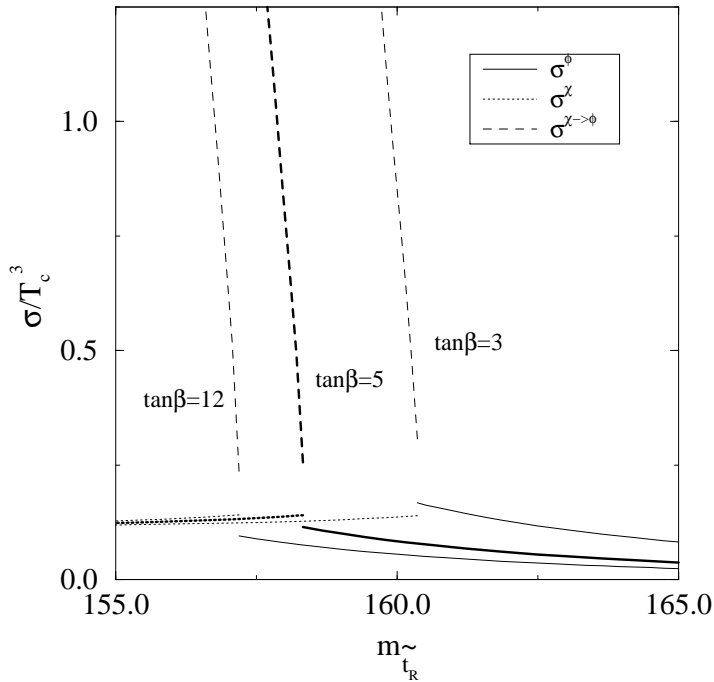}}

\figbottomspace

\caption[a]{
The surface tensions of the transitions 
as a function of $\mtR$ for $\tb=3,5,12$. The curves for the 
broken phase $\chi\to\phi$ transition represent an upper 
bound obtained as explained in the text.
Note that at $\mtRc$, 
$\sigma^{\chi\to\phi}=\sigma^{\phi}+\sigma^{\chi}$.
It is seen that for the broken phase transition, 
the surface tension becomes rapidly very large.
Metastable branches are not shown.}
\la{sigma}
\end{figure}

Results for the latent heat from~\nr{latent} 
as a function of $\tb, \mtR$ are shown in
Fig.~\ref{L}. 

To compute the surface tension~\nr{sigmaint} one has to find the path in
$(\phi,\chi)$ space which minimizes the integral.
Near $\mtR = \mtRc$ the mass parameters of the 3d theory are small.
Then the effective potential is dominated by the term $\gamma_3
\phi^2\chi^2/4$ in the tree level potential for $\phi,\chi \neq 0$
(see Fig.~\ref{3dfig}).
Therefore the path which minimizes (\ref{sigmaint}) 
for the transition $\chi\to \phi$
is close to the
rectangular path which first goes from $(0,\chi_{\rm min})$ to $(0,0)$
and then from $(0,0)$ to $(\phi_{\rm min},0)$. We have used this path
for estimating $\sigma^{\chi\to\phi}$ in
Fig.~\ref{sigma}.  The corresponding values are therefore an upper
limit for the surface tension.  At lower 
critical temperatures $T_c^{\chi\to\phi}$, corresponding
to smaller values of $\mtR$, the 3d mass parameters
decrease and our simple approximation becomes less reliable. We have
used different trial paths and compared the results for 
$\sigma^{\chi\to\phi}$ with
those from the rectangular path. From this we estimate the error as
at most 10\% for the parameter range shown in Fig.~\ref{sigma}.
This error estimate, of course, does not contain the effects
of higher-loop and non-perturbative contributions, which
might make $\sigma^{\chi\to\phi}$ much smaller. 

Based on Figs.~\ref{L}, \ref{sigma} 
and on Eqs.~\nr{nucl}, \nr{reheat}, one can say
the following:

1. The EW ($0\to\phi$) and CCB ($0\to\chi$) transitions
seem to proceed in the normal manner: there is at most 
a few percent supercooling, and the latent heat released
after the nucleation period reheats the matter
considerably. Especially
for larger Higgs masses in the EW phase transition, 
one might even reheat back to $T_c^\phi$.

2. For the broken phase transition ($\chi\to\phi$), 
the supercooling is considerably larger. For instance, 
at the point $\mtRc$ the latent heat is smaller
($L^{\chi\to\phi}=L^{\phi}-L^{\chi}$)
and the surface tension is larger
($\sigma^{\chi\to\phi}=\sigma^{\phi}+\sigma^{\chi}$)
than for the EW transition characterized by $L^\phi, \sigma^\phi$. 
Both effects increase the supercooling. The small 
supercooling formula~\nr{nucl} gives numbers of the order of 10\%, 
but the true supercooling might be even larger. Moreover, 
when one goes to smaller $\mtR$ the surface tension 
seems to grow more rapidly than the latent heat, so 
that one may end up in a situation where the lifetime
of the metastable CCB minimum is comparable with
cosmological time scales. This is clearly unacceptable. 
There is also no barrier temperature for
the broken phase transition~\cite{ccb}. Of course it
might happen that some other nucleation mechanism enters, 
in which case the transition could proceed after all;
or the non-perturbative surface tension might be
much smaller than our simple estimate.

3. If one happens to be very close to $\mtRc$, then 
things may become complicated. For instance, the supercooling
for the CCB transition is generally somewhat larger than
for the EW transition, so it is not excluded that at 
$\mtR\lsim \mtRc$ one supercools so much that the 
metastable branch of $T_c^\phi$ in Fig.~\ref{Tc} 
becomes relevant. Then the EW transition might take place
first. But after it has taken place, the CCB minimum is 
still the global one until $T=T_c^{\chi\to\phi}$, so that 
in principle two other transitions (first to the CCB minimum
and then back to the EW minimum at $T<T_c^{\chi\to\phi}$) 
could occur. However, this possibility clearly 
requires fine tuning.

\section{Discussion and conclusions}
\la{conclusions}

We have studied the electroweak phase transition in MSSM with $\mtR
< m_t$ at the 2-loop level of perturbation theory. It was found that
in principle there is the possibility that the EW phase transition
takes place in two stages. The first part, the colour breaking
transition, would delay and strengthen the transition to the standard
electroweak minimum. 
The sphaleron rate should not be significantly switched off in the
colour breaking minimum since the $U$-field is an SU(2) singlet, but
it would be very effectively switched off after the transition to the
standard electroweak minimum for Higgs masses up to 100 GeV  
(in the standard minimum normal electroweak
sphaleron estimates apply~\cite{moq}). Our
calculations were performed in a particular region of the parameter
space, with a relatively large CP-odd Higgs mass $m_A$ and a relatively
small left-handed third generation squark mass $m_Q \sim 300$ GeV, but
it appears that relaxing these assumptions does not immediately change
the qualitative pattern observed.

The problem with the two stage scenario is that
the parameter region in $\mtR$
allowing for a two stage transition is rather small, 
$\mtR\sim 155-160$ GeV. There
is always the danger that even though the CCB minimum 
becomes metastable, the first order transition 
to the standard EW minimum becomes so strong that
the lifetime of the metastable state is comparable
with cosmological time scales. 
Thus one may also interpret our result as a lower bound
$\mtR \gsim 155$ GeV for the right-handed stop mass.
This particular number is 
valid for small squark mixing parameters. 

Finally, it should be remembered that our
study was based on 2-loop perturbation theory. In fact,
already the large gauge and $\bmu$-dependence 
of our results indicates that
higher-loop perturbative (and possibly non-perturbative) 
effects could be 
much larger than in the SU(2)+Higgs model which
is relevant for the Standard Model and 
has been studied on the lattice.
If the right-handed stop mass
turns out to be such that the scenario
studied could be realized, then one should
probably consider lattice simulations in the 3d
SU(3)$\times$SU(2) effective theory.

\section*{Acknowledgements}

We are grateful to M. Carena, K. Kajantie, P.\ Overmann, 
M. Shaposhnikov and C.E.M. Wagner for discussions, 
and to A. Laser for collaboration during the initial
stages of this work. 
M.L was partially supported by the University of
Helsinki.

\appendix
\renewcommand{\thesection}{Appendix~\Alph{section}}
\renewcommand{\theequation}{\Alph{section}.\arabic{equation}}

\section{}
\la{app1}

In this Appendix, we discuss in some detail how the parameters 
in~\nr{Uthe} can be fixed at 1-loop (and partly 2-loop) level. 

\subsubsection*{$T=0$ parameters}

Since our main point is studying 3d IR phenomena and
since an accurate derivation of the 3d theory 
with the whole MSSM mass
spectrum included would be somewhat complicated, we will consider 
a rather simple special case here. The final effective theory~\nr{Uthe}
will be the same in many other cases as well, but the expressions
for the parameters will be different. We
consider the case of vanishing mixing in the squark mass
matrix and a heavy CP-odd Higgs particle ($m_A \gsim 300$ GeV)
so that only one Higgs doublet remains; 
this is the most favourable case for baryogenesis~\cite{beqz}.
Neither of these assumptions affects the form of the 
final effective theory. For instance, for two Higgs doublets
one can diagonalize
the Higgs mass matrix close to $T_c$ and integrate
the heavy doublet out~\cite{klrs1,ck,lo,ml}. We 
mostly neglect the U(1) gauge 
coupling, which should be a good approximation 
with respect to the other uncertainties in the calculation.
Finally, we assume that the only sparticles light enough to enter
the calculation are the squarks in the third generation. Adding
light fermionic degrees of freedom would again not change the
form of the final effective theory, only its 
parameters (see~\cite{lo} for the typical corrections arising). 

We shall for definiteness consider a scenario with 
a relatively light left-handed squark mass parameter
$m_Q \sim 300$ GeV. This is at the lower end of the
phenomenological constraints obtained 
with a realistic top mass~\cite{eqz,beqz}, and
at the same time still allows the high-temperature
expansion for $T\sim 100$ GeV 
(we also take the right-handed sbottom mass parameter to
be $m_D\sim 300$ GeV, although this has little effect). 
A much heavier $m_Q^2$ does not change the pattern
we are investigating qualitatively, but the formulas
to be used simplify as the $m_Q^2$-corrections
decouple.
The mass parameter $m_U^2$ is supposed to be 
relatively small, and its effects are discussed in the main text.

The most important renormalization effects
of the $T=0$ vacuum theory with respect to our calculation
concern the $\msbar$ mass parameters $m_H^2(\bmu)$, 
$m_U^2(\bmu)$
(we work throughout in the $\msbar$ scheme with the scale parameter $\bmu$). 
Calculating the pole Higgs mass $m_H$ to
1-loop order, one finds that $m_H^2(\bmu)$ can be expressed 
in terms of physical quantities as 
\be
m_H^2(\bmu) = -\frac{m_H^2}{2}+
\frac{3}{16\pi^2} h_t^2\ssb m_Q^2\biggl(
2\ln\frac{\bmu}{m_{\tilde{t}_L}}+1 \biggr), \la{mH}
\ee
where only the dominant terms are kept. For instance, 
there is another term with $m_Q^2 \to m_U^2$ and 
$m_{\tilde{t}_L} \to m_{\tilde{t}_R}$, 
but this is much less important 
for the parameter values we are considering.
For the relation between $\tb$ and $m_H$, we use the 
standard 1-loop formula containing the leading correction, 
\be
m_H^2 = m_Z^2\cos^2\! 2\beta +  
\frac{3 g_W^2}{8 \pi^2} \frac{m_t^4}{m_W^2}
\ln\frac{m_{\tilde{t}_R}m_{\tilde{t}_L}}{m_t^2}. \la{mHtb}
\ee
Here, in the absence of mixing, 
\be
m_{\tilde{t}_R}^2 = m_U^2+m_t^2,\quad 
m_{\tilde{t}_L}^2 = m_Q^2 + m_t^2 + \fr12 m_Z^2 \cbb. \la{mtL}
\ee
We take $m_t=170$ GeV so that 
$m_t^{\rm pole}\sim 175$ GeV according to 
$m_t^{\rm pole}\sim m_t(1+\frac{4\alpha_S}{3\pi})$. 
The bottom Yukawa coupling is neglected. 

Similarly, the dominant term in the expression for $m_U^2(\bmu)$
is
\be
m_U^2(\bmu) = m_U^2+
\frac{1}{8\pi^2} h_t^2 m_Q^2\biggl(
2\ln\frac{\bmu}{m_{\tilde{t}_L}}+1 \biggr), \la{mU}
\ee
where $m_U^2$ is expressed in terms of physical 
masses through~\nr{mtL}.

\subsubsection*{Step 1: dimensional reduction}

Next we go to finite temperature.
After integration over non-zero Matsubara frequencies, the effective
theory for the thermodynamics of the electroweak phase
transition contains only bosonic $n=0$ Matsubara 
modes~[5--11,32]. 
The form of the effective theory can be written down
immediately using 3d gauge invariance (only the terms 
coupled to $H, U, A^a_i, C^A_i$ are shown explicitly below):
\ba
L & = &
\fr14 F^a_{ij}F^a_{ij}+\fr14 G^A_{ij}G^A_{ij} 
+(D_i^w H)^\dagger(D_i^w H)+m_{H3}^2 H^\dagger H+
\lambda_{H3} (H^\dagger H)^2 \nn \\
& + & (D_i^s U^*)^\dagger(D_i^s U^*)+m_{U3}^2U^\dagger U+
\lambda_{U3} (U^\dagger U)^2
+ \gamma_3 H^\dagger H U^\dagger U \nn \\
& + & (D^{ws}_i Q)^\dagger(D^{ws}_i Q)+
(D^s_i D^*)^\dagger(D^s_i D^*) + 
m_{Q3}^2 Q_\alpha^\dagger Q_\alpha+
m_{D3}^2 D_\alpha^* D_\alpha \nn \\
& + & 
h_{t3}^2\Bigl(
Q^*_{i\alpha}U^*_{\alpha}Q_{i\beta}U_\beta+
\ssb\tilde{H}^{\dagger}Q_{\alpha}Q^\dagger_\alpha
\tilde{H}\Bigr) \nn \\
& + & 
\fr14 g_{W3}^2
Q^*_{i\alpha}Q_{j\alpha}
\Bigl[\ssb H^*_k H_l+\ccb \tilde{H}^{*}_k\tilde{H}_l\Bigr]
(2\delta_{il}\delta_{jk}-
\delta_{ij}\delta_{kl}) \nn \\
& + & 
\fr16 g_{S3}^2
U_\alpha U^*_\beta
\Bigl[D_\gamma D^*_\delta-Q^*_{j\gamma}Q_{j\delta}
\Bigr]
(3\delta_{\alpha\delta}\delta_{\beta\gamma}-
\delta_{\alpha\beta}\delta_{\gamma\delta}) \nn \\
& + & 
\fr12 (D_i^wA_0^a)^2+\fr12 m_{A_0}^2 A_0^aA_0^a +
\fr12 (D_i^s C_0^A)^2 + \fr12 m_{C_0}^2 C_0^AC_0^A \nn \\
& + & 
\fr14 g_{W3}^2 \hh A_0^aA_0^a + 
\fr14 g_{S3}^2 C_0^AC_0^B (U^*)^\dagger \lambda^A\lambda^B U^*,
\la{DRaction}
\ea
where
$D_i^wA_0^a=\partial_iA_0^a+g_{W3}\epsilon^{abc}A^b_iA^c_0$ and
correspondingly for $D_i^sC_0^A$, and $\tilde{H}=i\tau_2 H^*$.
We recall that, after trivial rescaling with $T$, the dimension
of bosonic fields in 3d is GeV$^{1/2}$ and that of
the quartic couplings is GeV.
In the interactions involving the fields 
$Q,D,A^a_0,C^A_0$ which are to be integrated out shortly, 
we take the parameters only at tree-level. 
What then remains to be discussed are
the values of the gauge couplings, the scalar couplings
$\lambda_{H3}, \lambda_{U3}, \gamma_3$,
and the mass parameters.

The gauge coupling in the dimensionally reduced 
SU($N$) gauge theory with $N_S$ scalars and
$N_f$ fermions is~\cite{klrs1,hl}
\be
g_3^2 = T g^2(\bmu) \biggl\{
1+\frac{g^2(\bmu)}{48\pi^2}\biggl[
(22 N -N_S) \frac{\Lb}{2}-4 N_f\frac{\Lf}{2}+N \biggr]\biggr\}, \la{g32}
\ee
where
\be
g^2(\bmu) = \frac{48\pi^2}{(22N-N_S-4N_f)\ln(\bmu/\Lambda)},
\ee 
and the standard
logarithmic corrections from bosonic and fermionic integrals are
\ba
L_b(\bmu) & = & 2 \ln\frac{\bmu e^{\gamma_E}}{4\pi T}\approx
2 \ln\frac{\bmu}{7.0555T}, \nn \\
L_f(\bmu) & = & 2 \ln\frac{\bmu e^{\gamma_E}}{\pi T}\approx
2 \ln\frac{\bmu}{1.7639T}. \la{Lb}
\ea
With the mass spectrum we are considering, for SU(3) the 
constants are $N=3$, $N_S\equiv N_S^s =4$ (four scalar triplets: $U$, $D$ and
the two components of $Q$), $N_f=6$. For SU(2)
we have $N=2$, 
$N_S\equiv N_S^w=4$ (one Higgs doublet and the three components of $Q$),
$N_f=6$ (three families, each with four chiral doublets). 

The coupling $g_3^2$ is independent of $\bmu$ to the order it
has been calculated, but it depends on the 
temperature. To make this dependence simple, we rewrite~\nr{g32} as
\be
g_3^2 = T\biggl[
g_0^2 + \frac{g_0^4}{48\pi^2}(22N-N_S-4N_f)\ln\frac{T_0}{T}
\biggr],
\ee
where $T_0\equiv 100$ GeV and $g_0^2$ is $g_3^2/T$ at 
the temperature $T_0$. For SU(3) with $\alpha_S(m_Z)\approx 0.12$
we get $g_{S0}^2 \sim 1.1$, and for SU(2) with 
$g_W(m_Z) \approx 2/3$ we get $g_{W0}^2 \sim 0.42$.
The effects of the U(1) gauge coupling are small 
and will hence be approximated by  $g' = 1/3$.
We denote $\hat{g}_{S3}^2 = g_{S3}^2/T$, 
$\hat{g}_{W3}^2 = g_{W3}^2/T$.

For the scalar self-coupling we only include the dominant 
$h_t^4$-correction, arising from the incomplete cancellation
of bosonic and fermionic integrals at finite temperature:
\ba
\frac{\lambda_{H3}}{T} 
& = &  \fr18 (g_W^2+g'^2) \cos^2\! 2\beta\; + 
\frac{3}{16\pi^2} h_t^4 \sin^4\!\beta\; \Bigl(\Lf-\Lb\Bigr) \nn \\
& = & 
\fr18 (g_W^2+g'^2) \cos^2\! 2\beta\;+
\frac{3}{4\pi^2} h_t^4 \sin^4\!\beta\;\ln 2.
\ea
For $\gamma_3$, $\lambda_{U3}$, it is more
difficult to give the dominant radiative corrections. If one calculates the
first corrections with the particle spectrum we are using, then, 
for instance,  one 
finds logarithmic terms in $\lambda_{U3}$ 
which are not the same as
for the gauge coupling $g_{S3}^2$ related directly to the gauge 
fields, although at tree-level $\lambda_{U3}\propto g_{S3}^2$. 
This is because supersymmetry has been broken by the
particle masses and would be restored only at a higher scale.  
However, the relative effect of these corrections is smaller
in $\gamma_3, \lambda_{U3}$ than in $\lambda_{H3}$, since 
the leading term is itself proportional to a large coupling.
Hence it is sufficient for the present purpose to use the
tree-level values
\be
\lambda_{U3}= \fr16 g_{S3}^2,\quad
\gamma_3 = h_t^2 \ssb T,
\ee
where $h_t^2\ssb = g_W^2 m_t^2/(2 m_W^2)$.

The Higgs sector mass parameter $m_{H3}^2$ gets modified
by the standard thermal screening terms (e.g. \cite{e}), 
and by a logarithmic 
$\Lb$-term cancelling the running in~\nr{mH}:
\ba
m_{H3}^2 & = & 
-\frac{m_H^2}{2}+\frac{3}{16\pi^2} h_t^2 \ssb m_Q^2
\biggl(2 \ln\frac{7.06 T}{m_{\tilde{t}_L}}+1\biggr) \nn \\
& + & \biggl( 
\frac{1}{16}(g_W^2+g'^2) \cos^2\! 2\beta\; +
\frac{3}{16} g_W^2 + \frac{1}{16}g'^2+
\fr34 h_t^2 \ssb
\biggr)T^2. \la{mmH3}
\ea
The logarithmic running is numerically very significant, 
since it is proportional to the large parameter $m_Q^2 \gg m_H^2$.
The scale at which the couplings in the screening part
are evaluated can only be fixed with a 2-loop calculation, 
but since it is established by the same $\Lb$ and $\Lf$ 
functions as for the gauge coupling in~\nr{g32}, we shall 
use the numerical values of $\hat{g}_{S3}^2$, $\hat{g}_{W3}^2$ for
$g_S^2, g_W^2$ everywhere in the finite temperature formulas.   

Similarly, $m_{U3}^2$ is
\be
m_{U3}^2 = 
m_U^2+\frac{1}{8\pi^2} h_t^2 m_Q^2
\biggl(2 \ln\frac{7.06 T}{m_{\tilde{t}_L}}+1\biggr) 
+ \biggl( 
\fr49 g_S^2+\fr16 h_t^2(1+\ssb)
\biggr)T^2. \la{mmU3}
\ee
Again the logarithmic running is numerically
significant.

For the other mass parameters we only include the leading
screening terms:
\ba
m_{Q3}^2 & = & m_Q^2
+\biggl(\fr14 g_W^2+\fr49 g_S^2+\frac{1}{12} h_t^2 (1+\ssb)
\biggr)T^2 , \la{mmQ3} \\
m_{D3}^2 & = & m_D^2
+\fr49 g_S^2 T^2, \la{mmD3} \\
m_{A_0}^2 & =  & g_W^2
\biggl(\fr23+\frac{N_f}{6}+\frac{N_S^w}{6}
\biggr)T^2, \la{ma0}\\
m_{C_0}^2 & =  & g_S^2
\biggl(1+\frac{N_f}{6}+\frac{N_S^s}{6}
\biggr)T^2, \la{mc0}
\ea 
where $N_f=6$ (see the explanation below~\nr{Lb}).
Note, in particular, that there are similar logarithmic 
runnings in~\nr{mmQ3}, \nr{mmD3}
as in~\nr{mmH3}. However, for $m_Q^2$ and $m_D^2$
these are not that important, as the parameters are themselves large. 

\subsubsection*{When is step 1 accurate?}

There are two basic requirements for the construction 
of the effective theory in~\nr{DRaction} to be accurate. 
First, the perturbative expansion for the parameters of
the effective theory should converge. This expansion is
free of IR-problems and thus proceeds just as perturbation theory
at zero temperature, except that 
the mass scale is $\sim 2 \pi T$. Thus there should be no problems
in the $\tb$ region considered. 
Second, the higher-order operators neglected in~\nr{DRaction}
should be insignificant. Such operators arise from the mass
hierarchy of the scales kept and integrated out, in other
words from the high-temperature expansion. Thus the effects
are small provided that $m^2/(2\pi T)^2 \ll 1$, where
$m$ symbolizes the mass scales in the Lagrangian~\nr{DRaction}.
This requirement is satisfied when the transition is not
exceedingly strong.

\subsubsection*{Step 2: heavy scale integrations}

To simplify~\nr{DRaction}, we will integrate
out the 3d scales which are heavy at the transition 
point, namely $Q, D, A_0, C_0$~\cite{ck,lo,ml}.
Afterwards, we make the replacement $U^* \to U$.
The result is the action in~\nr{Uthe}. 

The 1-loop relations between the parameters in~\nr{Uthe} 
and \nr{DRaction} are straightforward to derive
using the techniques in~\cite{klrs1,ck,lo,ml}. For
the gauge couplings, one gets
\ba
g_{W3}^{2(\rm new)} &  =  & 
g_{W3}^2 \biggl[
1-\frac{2 g_{W3}^2}{48\pi m_{A_0}}-\frac{3g_{W3}^2}{48\pi m_{Q3}}
\biggr], \\
g_{S3}^{2(\rm new)} &  =  & 
g_{S3}^2 \biggl[
1-\frac{3 g_{S3}^2}{48\pi m_{C_0}}-\frac{2g_{S3}^2}{48\pi m_{Q3}}
-\frac{g_{S3}^2}{48\pi m_{D3}}
\biggr].
\ea
Here the coefficient in the numerator of the first 
correction term is $N$, and in the subsequent terms 
it is the number of components of the corresponding fields. 

The mass parameters and couplings change to be 
\ba
m_{H3}^{2{({\rm new})}} & = & 
m_{H3}^2-\frac{3}{16\pi} g_{W3}^2 m_{A_0}
- \frac{3}{4\pi} h_{t3}^2\ssb m_{Q3}, \la{mmbH3} \\
m_{U3}^{2{({\rm new})}} & = & 
m_{U3}^2-\frac{1}{3\pi} g_{S3}^2 m_{C_0}
- \frac{1}{2\pi} h_{t3}^2 m_{Q3}, \la{mmbU3} \\
\lambda_{H3}^{({\rm new})} & = & 
\lambda_{H3} - \frac{3}{16}\frac{g_{W3}^4}{8 \pi m_{A_0}} \nn \\
& - & \frac{3}{16}\frac{1}{8 \pi m_{Q3}}
\Bigl(g_{W3}^4 \cos^2\! 2\beta+ 4 g_{W3}^2h_{t3}^2
\cos\! 2\beta\ssb+ 8 h_{t3}^4\sin^4\! \beta\Bigr), \\
\lambda_{U3}^{({\rm new})} & = & 
\lambda_{U3} - \frac{13}{36}\frac{g_{S3}^4}{8 \pi m_{C_0}}-
\frac{1}{12}\frac{g_{S3}^4}{8 \pi m_{D3}} \nn \\
& - & \frac{1}{6}\frac{1}{8 \pi m_{Q3}}
\Bigl(g_{S3}^4 - 4 g_{S3}^2h_{t3}^2
+ 6 h_{t3}^4\Bigr), \\
\gamma_3^{({\rm new})} & = & 
\gamma_3 - \frac{h_{t3}^4\ssb}{8 \pi m_{Q3}}.
\ea

\subsubsection*{When is step 2 accurate?}

The requirements for the integration in step 2 to be accurate are the same
as for the dimensional reduction step. First, 
the perturbative expansion for the parameters should converge. 
The expansion parameters were estimated in~\cite{ml}, and 
are roughly 
\be
\frac{g_{W3}^2}{4\pi m_{A_0}}, \quad
\frac{g_{S3}^2}{\pi m_{C_0}} \la{a0exp}
\ee
for the integration over $A_0$, $C_0$, and  
\be
\frac{g_{S3}^2}{\pi m_{Q3}}, \quad
\frac{h_{t3}^2}{\pi m_{Q3}} \la{expU3}
\ee
for the integration over $Q$, $D$ (for $D$, $m_{Q3}\to m_{D3}$).
Evaluating the numerical values, one can see that
when the 1-loop corrections are kept, the 
neglected 2-loop terms are small. Still, if the critical temperature
is needed very precisely, the 2-loop corrections to the 
mass parameters would be needed (see below).

The second requirement is that neglected
higher-order operators should give small contributions.
Such operators arise in particular from the expansion
of the masses of the fields in~\nr{DRaction} in the background 
$\langle H^\dagger H\rangle = \phi^2/2$, 
$\langle U^\dagger U\rangle = \chi^2/2$, in terms of the 
background fields. 
Typical 1-loop terms to be expanded are
\ba
& & -\frac{1}{4\pi}\Bigl( m_{A_0}^2+g_W^2\phi^2/4\Bigr)^{3/2}, \quad
-\frac{1}{3\pi}\Bigl( m_{C_0}^2+g_S^2\chi^2/4\Bigr)^{3/2}, \nn \\
& & -\frac{1}{6\pi}
\Bigl[
m_{Q3}^2+(4 h_t^2 \ssb+g_W^2 \cbb)\phi^2/8+
(3 h_t^2-g_S^2)\chi^2/6 \Bigr]^{3/2}, \nn \\
& & -\frac{1}{6 \pi} \Bigl(
m_{D3}^2 + g_S^2 \chi^2/6
\Bigr)^{3/2}. \la{errors}
\ea
The higher-order operators arise from the fourth terms
in the expansions. In~\nr{errors}, 
the couplings and fields are for simplicity in 4d units.
Taking the actual parameter values one can see
that the expansion is valid for 
$\phi/T \lsim 3$, $\chi/T \lsim 3$.

\subsubsection*{2-loop mass parameters}

So far we have discussed the mass parameters 
in~\nr{Uthe} at the 1-loop level ($\sim g_S^2T^2$). 
However, we are calculating the effective potential in 3d 
at the 2-loop level, and then the mass parameters get 
renormalized. In fact, the renormalized parts 
of the mass parameters in the $\msbar$ scheme 
turn out to be of the form  
\ba
m_{H3}^2(\bmu) \!\! & = & \!\! 
m_{H3}^2 +\pf \Bigl(
\frac{51}{16}g_{W3}^4+9 \lambda_{H3}g_{W3}^2-12\lambda_{H3}^2-
3\gamma_3^2+8 g_{S3}^2 \gamma_3
\Bigr)\ln \frac{\Lambda_{H3}}{\bmu}, \hspace*{0.8cm} 
\la{mmH3mu}\\
m_{U3}^2(\bmu) \!\! & = & \!\! 
m_{U3}^2 +\pf \Bigl(
8 g_{S3}^4+\frac{64}{3} \lambda_{U3}g_{S3}^2-16\lambda_{U3}^2-
2\gamma_3^2+3 g_{W3}^2 \gamma_3
\Bigr)\ln \frac{\Lambda_{U3}}{\bmu}, \la{mmU3mu}
\ea
where $m_{H3}^2, m_{U3}^2$ are the 1-loop expressions
in~\nr{mmbH3}, \nr{mmbU3}.
To fix $\Lambda_{H3}$, $\Lambda_{U3}$ 
would require a 2-loop derivation of the mass parameters 
(this is essentially a 2-loop calculation of the effective potential in 4d;
this is the {\it only} place where a 2-loop calculation in 4d gives
information not contained in a 2-loop calculation in 3d~\cite{fkrs1,klrs1}).
We will not make the 2-loop
derivation here but shall instead take the
order of magnitude estimate 
\be
\Lambda_{H3} \sim \Lambda_{U3} \sim 7 T.
\ee
This estimate arises from the typical mass scales of
integrated-out degrees of freedom and can be confirmed in 
the case of the Standard Model, where a 2-loop derivation
of the mass parameters has been made~\cite{klrs1}.
The uncertainty from the choice
of $\Lambda$'s affects the critical temperatures, 
but does not affect the conclusions concerning the 2-loop effects
on dimensionless ratios like 
$\phi/T_c$, $\chi/T_c$, $\hat{L}= L/T_c^4$, 
$\hat{\sigma}= \sigma/T_c^3$, where $L$ is the 
latent heat and $\sigma$ is the surface tension. 
The estimated uncertainty in $T_c$ is on the level 
of few ($\lsim 2-3$) percent.

This concludes the derivation of the parameters in
the effective theory~\nr{Uthe}.

\section{}
\la{app2}

In this Appendix, we present some details of the 
calculation of the 2-loop effective potential 
$V(\phi,\chi)$ in the theory defined by~\nr{Uthe}
(with SU(3) replaced by SU($N$)).
To compactify the expressions, we denote $g_{W3}\to g_W,
g_{S3}\to g_S,  \lambda_{H3}\to \lambda_H, 
\lambda_{U3} \to \lambda_U$.

To derive $V(\phi,\chi)$, one has to shift the 
fields $H$ and $U$ by the constants
$\hat{H}$ and $\hat{U}$ and then to calculate 
all the vacuum diagrams in the shifted theory~\cite{jw}.
Here we are using the notation 
\be
\hat{H}^\dagger\hat{H}=\frac{\phi^2}{2},\quad
\hat{U}^\dagger\hat{U}=\frac{\chi^2}{2}.
\ee
For simplicity, we shall assume the shifts 
$\hat{H}$ and $\hat{U}$ to be real but otherwise
arbitrary (the assumption of realness 
enters in the scalar propagators). 
We use the background field gauge fixing conditions
(the applicability of these has been discussed in~\cite{kls1}), 
\ba
G^a & = & \partial_i A_i^a + \frac{i}{2}\xi g_W 
\Bigl(\hat{H}^\dagger\tau^a H-\hd \tau^a\hat{H}\Bigr), \\
G^A & = & \partial_i C_i^A + \frac{i}{2}\zeta g_S 
\Bigl(\hat{U}^\dagger\lambda^A U-U^\dagger \lambda^A\hat{U}\Bigr),
\ea
where $\tau^a$ are the Pauli matrices and $\lambda^A$ are
SU($N$) generators, normalized such that 
$\mathop{\rm Tr}\lambda^A\lambda^B = 2\delta^{AB}$.
The terms added to the action are 
\be
\frac{1}{2\xi}G^aG^a,\quad \frac{1}{2\zeta}G^AG^A, 
\ee 
supplemented by the corresponding ghost terms.
The limit $\xi\to 0$, 
$\zeta\to 0$, which we use for most of the numerical 
evaluations, gives the usual Landau gauge.

The shift of $H$, $U$ is made by writing the original fields as
\ba
H & = & \hat{H}+\tilde{H},\quad \tilde{H}_i = \frac{1}{\sqrt{2}}
(h_i+i\pi_i),\quad i=1,2, \nonumber \\
U & = & \hat{U}+\tilde{U},\quad \tilde{U}_\alpha = \frac{1}{\sqrt{2}}
(u_\alpha+i\omega_\alpha),\quad \alpha=1,\ldots ,N. \la{fields}
\ea
The gauge fields are not shifted. 
In the shifted theory, the mass spectrum is such that 
the (non-vanishing) gauge bosons masses are
\be
m_W^2=\fr14 g_W^2\phi^2,\quad
m_G^2=\fr14 g_S^2\chi^2,\quad
\overline{m}_G^2=\frac{2(N-1)}{N}m_G^2,
\ee
and the Goldstone boson masses are
\ba
m_\pi^2 & = & m_{H3}^2+\lambda_H\phi^2+\fr12\gamma_3\chi^2+\xi m_W^2,
\nonumber \\
m_\omega^2 & = & m_{U3}^2+\lambda_U\chi^2+\fr12\gamma_3\phi^2+\zeta m_G^2,
\nonumber \\
\overline{m}_\omega^2 
& = & m_{U3}^2+\lambda_U\chi^2+\fr12\gamma_3\phi^2+\zeta \overline{m}_G^2.
\ea
In the Higgs sector there is mixing, so that the eigenstates are
\be
m_{\pm}^2=\fr12\biggl[
m_h^2+m_u^2 \pm
\sqrt{(m_h^2-m_u^2)^2+4 \gamma_3^2 \phi^2 \chi^2}
\biggr],
\ee
where
\be
m_h^2=m_{H3}^2+3\lambda_H\phi^2+\fr12\gamma_3\chi^2, \quad
m_u^2=m_{U3}^2+3\lambda_U\chi^2+\fr12\gamma_3\phi^2.
\ee
We denote the mixing angles by
\be
s^2=\frac{m_+^2-m_h^2}{m_+^2-m_-^2}, \quad
c^2=\frac{m_h^2-m_-^2}{m_+^2-m_-^2}, \quad 
cs = -\frac{\gamma_3\phi\chi}{m_+^2-m_-^2}.
\ee

The propagators in the shifted theory are as follows. Let
\ba
T_{\alpha\beta}=\delta_{\alpha\beta}-
\frac{\hat{U}_\alpha\hat{U}_\beta}{\hat{U}^\dagger\hat{U}},\quad
L_{\alpha\beta}=
\frac{\hat{U}_\alpha\hat{U}_\beta}{\hat{U}^\dagger\hat{U}},
\ea
and correspondingly for the $\hat{H}$-direction. Then 
\ba
\langle h_i h_j\rangle &  = &  \frac{T_{ij}}{k^2+m_\pi^2}+
\biggl( \frac{c^2}{k^2+m_+^2}+\frac{s^2}{k^2+m_-^2}
\biggr) L_{ij}, \\
\langle\pi_i\pi_j\rangle &  = &  \frac{\delta_{ij}}{k^2+m_\pi^2}, \\
\langle u_\alpha u_\beta\rangle &  = &  \frac{T_{\alpha\beta}}{k^2+
m_\omega^2}+
\biggl( \frac{s^2}{k^2+m_+^2}+\frac{c^2}{k^2+m_-^2}
\biggr) L_{\alpha\beta}, \\
\langle\omega_\alpha\omega_\beta\rangle &  = &  \frac{T_{\alpha\beta}}{k^2+
m_\omega^2}+ \frac{L_{\alpha\beta}}{k^2+\overline{m}_\omega^2}, \\
\langle h_i u_\alpha\rangle & = & 
\frac{\hat{H}_i\hat{U}_\alpha}{\sqrt{\hat{H}^\dagger\hat{H}
\hat{U}^\dagger\hat{U}}}\biggl(
\frac{cs}{k^2+m_-^2}-\frac{cs}{k^2+m_+^2}\biggr). \la{xprop}
\ea
For the vector and ghost propagators, we define the 
projection operators
\ba
P_1^{AB} & = & 
\delta^{AB}+\fr12\biggl(\frac{N-2}{N-1}\biggr)
\frac{\hat{U}^\dagger\lambda^A\hat{U}}{\hat{U}^\dagger\hat{U}}
\frac{\hat{U}^\dagger\lambda^B\hat{U}}{\hat{U}^\dagger\hat{U}}
-\fr12
\frac{\hat{U}^\dagger\lambda^{\{A}\lambda^{B\}}
\hat{U}}{\hat{U}^\dagger\hat{U}}, \\
P_2^{AB} & = & 
-\frac{\hat{U}^\dagger\lambda^A\hat{U}}{\hat{U}^\dagger\hat{U}}
\frac{\hat{U}^\dagger\lambda^B\hat{U}}{\hat{U}^\dagger\hat{U}}
+\fr12
\frac{\hat{U}^\dagger\lambda^{\{A}\lambda^{B\}}
\hat{U}}{\hat{U}^\dagger\hat{U}}, \\
P_3^{AB} & = & 
\fr12
\frac{N}{N-1}\frac{\hat{U}^\dagger\lambda^A\hat{U}}{\hat{U}^\dagger\hat{U}}
\frac{\hat{U}^\dagger\lambda^B\hat{U}}{\hat{U}^\dagger\hat{U}},
\ea
based on $\lambda^A_{\alpha\beta}
\lambda^A_{\gamma\delta}=
2(\delta_{\alpha\delta}\delta_{\beta\gamma}-
\delta_{\alpha\beta}\delta_{\gamma\delta}/N)$ 
and satisfying $\sum_{n=1}^3 P_n^{AB}=\delta^{AB}$.
Then 
\ba
\langle C^A_i C^B_j\rangle & = & 
\biggl(\delta_{ij}-\frac{k_ik_i}{k^2}\biggr)
\biggl(\frac{P_1^{AB}}{k^2}+\frac{P_2^{AB}}{k^2+m_G^2}+
\frac{P_3^{AB}}{k^2+\overline{m}_G^2}
\biggr) \nonumber \\
& & +\zeta\frac{k_ik_i}{k^2}
\biggl(\frac{P_1^{AB}}{k^2}+\frac{P_2^{AB}}{k^2+\zeta m_G^2}+
\frac{P_3^{AB}}{k^2+\zeta\overline{m}_G^2}
\biggr), \la{CAprop} \\
\langle \overline{c}^A c^B\rangle & = & 
-\biggl(\frac{P_1^{AB}}{k^2}+\frac{P_2^{AB}}{k^2+\zeta m_G^2}+
\frac{P_3^{AB}}{k^2+\zeta\overline{m}_G^2}
\biggr). \la{caprop}
\ea
Here $\overline{c}^A, c^B$ are the ghost fields.
For SU(2), $P_1^{AB}=0$ and $m_G^2, \overline{m}_G^2 \to m_W^2$;
Eqs.~\nr{CAprop}, \nr{caprop} then
reduce to the standard isospin diagonal ($\delta^{ab}$) propagators.

The 1-loop potential of the theory can be easily calculated. 
Using 
$P_2^{AA}=2(N-1)$, $P_3^{AA}=1$,
the result is 
\ba
V_{\rm 1-l}(\phi,\chi) & = &
\fr12 m_{H3}^2\phi^2+\fr14\lambda_H\phi^4+
\fr12 m_{U3}^2\chi^2+\fr14\lambda_U\chi^4+
\fr14\gamma_3\phi^2\chi^2 \nonumber \\
& - & \frac{1}{12\pi}\biggl[ 
m_+^3+m_-^3+3m_\pi^3+2(N-1)m_\omega^3+\overline{m}_\omega^3
\nonumber \\ 
& + & 3(2-\xi^{3/2}) m_W^3+
2(N-1) (2-\zeta^{3/2}) m_G^3+
(2-\zeta^{3/2})\overline{m}_G^3 \biggr]. 
\ea

For the 2-loop potential, we need a number of integrals. 
Let
\ba
S(k^2,m^2) & = &  \frac{1}{k^2+m^2}, \\
V^\zeta_{ij}(k^2,M^2) & = & 
\frac{\delta_{ij}-k_ik_j/k^2}{k^2+M^2}+
\zeta\frac{k_ik_j/k^2}{k^2+\zeta M^2}, \\
F_{ijk}(p,q,r) & = & 
\delta_{ik}(p_j-r_j)+\delta_{kj}(r_i-q_i)+\delta_{ji}(q_k-p_k).
\ea
Then we define the following integrals~\cite{ae,fkrs1}, in terms
of which the 2-loop potential can be written:
\ba
D^\zeta_{\rm VVV}(M_1^2,M_2^2,M_3^2) & = & \!\!
\int\!dp\,dq\,
V^\zeta_{il}(p^2,M_1^2) 
V^\zeta_{jm}(q^2,M_2^2) 
V^\zeta_{kn}(r^2,M_3^2) \nonumber \\
& & \quad\quad \times F_{ijk}(p,q,r)F_{lmn}(p,q,r), \la{dvvv} \\
D^\zeta_{\rm VVS}(M_1^2,M_2^2,m^2) & = & \!\! 
\int\!dp\,dq\,
V^\zeta_{il}(p^2,M_1^2) V^\zeta_{jm}(q^2,M_2^2) S(r^2,m^2) 
4\delta_{ij}\delta_{lm}, \\
D^\zeta_{\rm SSV}(m_1^2,m_2^2,M^2)  & = & \!\!  
\int\!dp\,dq\,
S(p^2,m_1^2) S (q^2,m_2^2) V^\zeta_{ij}(r^2,M^2) 
(p_i-q_i)(p_j-q_j),\hspace*{1.0cm} \\
D^\zeta_{\rm GGV}(m_1^2,m_2^2,M^2) & = & \!\!  
\int\!dp\,dq\,
S(p^2,m_1^2) S (q^2,m_2^2) V^\zeta_{ij}(r^2,M^2) 
(-p_iq_j), \la{dggv} \\
H(m_1^2,m_2^2,m_3^2) & = &  \!\!
\int\!dp\,dq\,
S(p^2,m_1^2) S (q^2,m_2^2) S(r^2,m_3^2), \la{h} \\
I^\zeta_V(M^2) & = & \!\!
\int\!dp\,
V^\zeta_{ij}(p^2,M^2)\delta_{ij} = 
(d-1) I_S(M^2)+\zeta I_S(\zeta M^2), \\
I_S(m^2) & = & \!\!
\int\!dp\, S(p^2,m^2), \la{IS}
\ea
where $r=-p-q$, $d$ is the space dimension, and
\be
dp = \frac{d^dp}{(2\pi )^d}.
\ee

All the $D$-integrals
in~\nr{dvvv}--\nr{dggv}
can be reduced to combinations
of $H$- and $I_S$-functions
defined in~\nr{h}, \nr{IS}. In the Landau gauge, this reduction
can be found in~\cite{ae,fkrs1}, 
and in the general gauge for mass combinations
relevant for SU(2) in~\cite{kls1} (note that in~\cite{fkrs1},
$D_{\rm VVV}$ is $-1/4$ and 
$D_{\rm VVS}$ is $1/4$ of our corresponding expression).
In the general gauge
with mass combinations relevant for SU($N$), 
we have done the reduction using 
symbolic manipulation packages. We do not present the final 
formulas here since some of them are rather lengthy. As an example, 
let us give the most complicated Landau gauge result
needed. For $d=3-2\epsilon$, $\zeta =0$, 
\ba
D^{\zeta=0}_{\rm VVV}(M^2,M^2,m^2) & = &
\frac{1}{16\pi^2}\biggl(
\frac{23}{3} M^2-m^2+\frac{11}{6}m M +\frac{5}{2}\frac{m^3}{M}-
\fr12 \frac{M^3}{m}-\fr14 \frac{m^4}{M^2}
\biggr) \nonumber \\
& + & H(m^2,M^2,M^2)
\biggl(
-8 M^2-10 m^2+2 \frac{m^4}{M^2}+\fr14 \frac{m^6}{M^4}
\biggr)\nonumber \\
& + & H(m^2,M^2,0)
\biggl(
-2 M^2+5 m^2-2 \frac{m^4}{M^2}-\fr12 \frac{m^6}{M^4}-
\fr12 \frac{M^4}{m^2} \biggr) \nonumber \\
& + & H(m^2,0,0)\frac{m^6}{4 M^4} + 
H(M^2,0,0) \frac{M^4}{2 m^2}, \la{xi0vvv}
\ea
where we have used that
\be
I_S(m^2) =  -\frac{m}{4\pi}. \la{intI}
\ee

For the general gauge, let us give just the 
divergent parts of the $D$-integrals:
\ba
D^\zeta_{\rm VVV}(M_1^2,M_2^2,M_3^2) & \to & 
\frac{1}{16\pi^2}\frac{1}{4\epsilon}
\biggl[-(5+\zeta)(M_1^2+M_2^2+M_3^2)\biggr], \la{divvvv}\\
D^\zeta_{\rm VVS}(M_1^2,M_2^2,m^2) & \to & 
\frac{1}{16\pi^2}\frac{1}{4\epsilon}
\biggl[6+4\zeta + 2\zeta^2\biggr], \\
D^\zeta_{\rm SSV}(m_1^2,m_2^2,M^2) & \to & 
\frac{1}{16\pi^2}\frac{1}{4\epsilon}
\biggl[M^2-2m_1^2-2m_2^2\biggr], \\
D^\zeta_{\rm GGV}(m_1^2,m_2^2,M^2) & \to & 
\frac{1}{16\pi^2}\frac{1}{4\epsilon}
\biggl[
\fr14 M^2(1+\zeta^2)-
\fr12(m_1^2+m_2^2)\biggr]. \la{divggv}
\ea
These arise from the 3d $H$-integral~\cite{fkrs1} 
defined in~\nr{h}, 
\be
H(m_1^2,m_2^2,m_3^2) = 
\frac{1}{16\pi^2}\biggl(
\frac{1}{4\epsilon}+\ln\frac{\bmu}{m_1+m_2+m_3}+\fr12\biggr).
\la{intH}
\ee

Using the $D$-integrals, one can write
down the 2-loop graphs. Let us start with the nonabelian
graphs, which we give for the SU($N$) sector; the SU(2)
sector is a special case of this. 
There are the sunset graphs (PPP) and the figure-8 graphs (PP)
(see, e.g., \cite{kls1,ae}),
containing vectors (P$\to$V), scalars (P$\to$S) and ghosts (P$\to$G):
\ba
({\rm VVV}) & = & 
-g_S^2 \frac{N}{12}
\biggl[
(N-1)(N-2) D^\zeta_{\rm VVV}(0,0,0)+
3(N-2) D^\zeta_{\rm VVV}(m_G^2,m_G^2,0) \nonumber \\
& & + 3 D^\zeta_{\rm VVV}(m_G^2,m_G^2,\overline{m}_G^2)
\biggr], \la{first} \\
({\rm GGV}) & = & 
g_S^2 \frac{N}{2}
\biggl[
(N-1)(N-2) D^\zeta_{\rm GGV}(0,0,0)+
2(N-2) D^\zeta_{\rm GGV}(0,\zeta m_G^2,m_G^2)  \nonumber \\
& & +(N-2) D^\zeta_{\rm GGV} (\zeta m_G^2,\zeta m_G^2,0)+
2 D^\zeta_{\rm GGV} (\zeta m_G^2,\zeta \overline{m}_G^2,m_G^2) \nonumber \\
& & + D^\zeta_{\rm GGV}(\zeta m_G^2,\zeta m_G^2,\overline{m}_G^2)
\biggr], \\
({\rm VVS}) & = & 
-g_S^2 \frac{m_G^2}{8}
\biggl[
(N-1) D^\zeta_{\rm VVS}(m_G^2,m_G^2,m_u^2)+
2\frac{(N-1)^2}{N^2} D^\zeta_{\rm VVS}(\overline{m}_G^2,
\overline{m}_G^2,m_u^2) \nonumber  \\
& & +N(N-2) D^\zeta_{\rm VVS} (0,m_G^2,m_\omega^2)+
\frac{(N-2)^2}{N}
D^\zeta_{\rm VVS} (m_G^2,\overline{m}_G^2,m_\omega^2)
\biggr], \\
({\rm SSV}) & = & 
-g_S^2 \frac{1}{4}
\biggl[
(N-1) D^\zeta_{\rm SSV}(m_\omega^2,\overline{m}_\omega^2,m_G^2)+
(N-1) D^\zeta_{\rm SSV}(m_\omega^2,m_u^2,m_G^2) \nonumber \\
& & +\frac{N-1}{N} D^\zeta_{\rm SSV} 
(\overline{m}_\omega^2,m_u^2,\overline{m}_G^2)+
\frac{1}{N} D^\zeta_{\rm SSV} (m_\omega^2,m_\omega^2,\overline{m}_G^2)
\nonumber \\
& & + N(N-2)D^\zeta_{\rm SSV} (m_\omega^2,m_\omega^2,0)
\biggr], \\
({\rm GGS}) & = & 
-\zeta^2 g_S^2 m_G^2\frac{N-1}{4}
\biggl[
H(\zeta m_G^2,\zeta m_G^2,\overline{m}_\omega^2)-
H(\zeta m_G^2,\zeta m_G^2,m_u^2) \nonumber \\
& & +\fr4N H (\zeta m_G^2,\zeta \overline{m}_G^2,m_\omega^2)-
\frac{2(N-1)}{N^2}
H (\zeta \overline{m}_G^2,\zeta \overline{m}_G^2,m_u^2)
\biggr], \\
({\rm VV}) & = & 
g_S^2 \frac{N}{4}\frac{d-1}{d}
\biggl[
(N-1)(N-2) I^\zeta_V(0) I^\zeta_V(0)+
2(N-2) I^\zeta_V(0) I^\zeta_V (m_G^2) \nonumber \\
& & + 2 I^\zeta_V(m_G^2)I^\zeta_V(\overline{m}_G^2)
+ (N-1) I^\zeta_V(m_G^2) I^\zeta_V(m_G^2)
\biggr], \\
({\rm SV}) & = & 
g_S^2 \fr14
\biggl\{
2 N (N-2) I^\zeta_V(0) I_S(m_\omega^2) \nonumber \\
& & +
(N-1) I^\zeta_V(m_G^2)\Bigl[ 
2 I_S(m_\omega^2) + I_S(\overline{m}_\omega^2) +
I_S(m_u^2) \Bigr] \nonumber \\
& & + \fr1N I^\zeta_V(\overline{m}_G^2)
\Bigl[ 
2 I_S(m_\omega^2) + (N-1) I_S(\overline{m}_\omega^2) +
(N-1) I_S(m_u^2) \Bigr]
\biggr\}. \la{last}
\ea
The integrals with all arguments zero 
vanish in the $\msbar$ scheme (in a general scheme
they give vacuum terms). We have also used
the conventions
\ba
f(\ldots,m_u^2,\ldots) & \equiv & 
s^2 f(\ldots,m_+^2,\ldots)+
c^2 f(\ldots,m_-^2,\ldots), \la{fmu}\\
f(\ldots,m_h^2,\ldots) & \equiv & 
c^2 f(\ldots,m_+^2,\ldots)+
s^2 f(\ldots,m_-^2,\ldots), \\
f(\ldots,m_x^2,\ldots) & \equiv & 
cs \Bigl[f(\ldots,m_-^2,\ldots)-
f(\ldots,m_+^2,\ldots) \Bigr]. \la{fmx}
\ea
The last one is needed below.

It is a useful check of~\nr{first}--\nr{last}
to sum up the divergent parts using~\nr{divvvv}--\nr{divggv}, 
and to verify that the result is gauge-independent. Indeed, 
the sum is 
\be
\frac{1}{16\pi^2}\frac{1}{4\epsilon}
(N^2-1)\biggl[
g_S^2 m_{U3}^2 + \frac{\phi^2}{2} g_S^2\gamma_3 
+\frac{\chi^2}{2}
\biggl(
g_S^4 \frac{4N^2-N+3}{4N^2}+2 \lambda_U g_S^2\frac{N+1}{N}
\biggr)
\biggr]. \la{divergence}
\ee
The field independent part of this expression 
can be cancelled with a vacuum 
counterterm, and the rest is cancelled 
by the mass counterterms (see below).

What remains is to write down the scalar graphs. The integrals appearing
are simple, but the result is complicated by the mixing of the 
two Higgs fields. To simplify the formulas, we employ~\nr{fmu}--\nr{fmx}.
A scalar line involving the fields $h, \pi$
is denoted by $H$, a line involving $u, \omega$
is denoted by $U$, and a line involving the 
mixed propagator~\nr{xprop} is denoted by $X$. 
We also leave out the subscript from the
integral $I_S$ defined in~\nr{IS}.
Then the results are:
\newcommand{\mmu}{m_u^2}
\newcommand{\mmh}{m_h^2}
\newcommand{\mmx}{m_x^2}
\newcommand{\mmpi}{m_\pi^2}
\newcommand{\mmom}{m_\omega^2}
\newcommand{\mmomb}{\overline{m}_\omega^2}
\ba
({\rm UUU}) \!\! & = & \!\!
-\lambda_U^2\chi^2
\Bigl[
3 H(m_u^2,m_u^2,m_u^2)+
2 (N-1) H(m_u^2,m_\omega^2,m_\omega^2) \nonumber \\
& & \quad +
H(m_u^2,\overline{m}_\omega^2,\overline{m}_\omega^2)
\Bigr], \\
({\rm HHH}) \!\! & = & \!\!
-\lambda_H^2\phi^2
\Bigl[
3 H(m_h^2,m_h^2,m_h^2)+
3 H(m_h^2,m_\pi^2,m_\pi^2)
\Bigr], \\
({\rm HUU}) \!\! & = & \!\!
-\fr14\gamma_3^2\phi^2
\Bigl[
H(\mmh,\mmu,\mmu)+
2(N-1) H(\mmh,\mmom,\mmom) \nonumber \\
& & \quad +H(\mmh,\mmomb,\mmomb)
\Bigr], \\
({\rm UHH}) \!\! & = & \!\!
-\fr14\gamma_3^2\chi^2
\Bigl[
H(\mmu,\mmh,\mmh)+
3 H(\mmu,\mmpi,\mmpi)
\Bigr], \\
({\rm UXX}) \!\! & = & \!\!
-\Bigl(\fr12\gamma_3^2\phi^2
+3\lambda_U\gamma_3\chi^2\Bigr)
H(\mmu,\mmx,\mmx), \\
({\rm HXX}) \!\! & = & \!\!
-\Bigl(\fr12\gamma_3^2\chi^2
+3\lambda_H\gamma_3\phi^2
\Bigr)
H(\mmh,\mmx,\mmx), \\
({\rm UUX}) \!\! & = & \!\!
-\lambda_U\gamma_3\phi\chi
\Bigl[
3H(\mmu,\mmu,\mmx)+
2(N-1)H(\mmom,\mmom,\mmx) \nonumber \\
& & \quad +
H(\mmomb,\mmomb,\mmx)
\Bigr], \\
({\rm HHX}) \!\! & = & \!\!
-\lambda_H\gamma_3\phi\chi
\Bigl[
3H(\mmh,\mmh,\mmx)+
3H(\mmpi,\mmpi,\mmx)
\Bigr], \\
({\rm UHX}) \!\! & = & \!\!
-\gamma_3^2\phi\chi H(\mmu,\mmh,\mmx), \\
({\rm XXX}) \!\! & = & \!\!
-\Bigl(\fr12\gamma_3^2
+6\lambda_U\lambda_H\Bigr)\phi\chi
H(\mmx,\mmx,\mmx), \\
({\rm UU})  \!\! & = & \!\!
\fr14\lambda_U
\Bigl[
3I(\mmu)I(\mmu)
+4(N-1)I(\mmu) I(\mmom)
+2 I(\mmu) I(\mmomb) \nonumber \\
& & \!\! +4N(N-1)I(\mmom) I(\mmom)
+4(N -1)I(\mmom) I(\mmomb)
+3I(\mmomb) I(\mmomb)
\Bigr], \hspace*{0.8cm} \\
({\rm HH}) \!\! & = & \!\!
\fr14\lambda_H
\Bigl[
3 I(\mmh)I(\mmh)+
6 I(\mmh)I(\mmpi)+
15 I(\mmpi)I(\mmpi)
\Bigr], \\
({\rm UH}) \!\! & = & \!\!
\fr14\gamma_3 
\Bigl[I(\mmu)+2(N-1)I(\mmom)+I(\mmomb)\Bigr]
\Bigl[I(\mmh)+3 I(\mmpi)\Bigr], \\
({\rm XX}) \!\! & = & \!\!
\fr12\gamma_3 I(\mmx)I(\mmx). 
\ea
The complete 2-loop contribution to $V(\phi,\chi)$ is obtained
by summing these scalar graphs together with the nonabelian 
graphs in~\nr{first}--\nr{last}, evaluated both for SU(2) and SU($N$).

The divergent part of the sum of the scalar graphs is
\be
-\frac{1}{16\pi^2}\frac{1}{4\epsilon}
\biggl[
\frac{\chi^2}{2}
\biggl(4(N+1)\lambda_U^2+2\gamma_3^2\biggr)
+\frac{\phi^2}{2}
\biggl(12 \lambda_H^2+N\gamma_3^2\biggr)
\biggr].
\ee
When one sums this (for $N=3$) with~\nr{divergence}, 
evaluated both for SU(2) and SU(3), one gets
exactly the divergences which after cancellation 
by the mass counterterms, leave for the renormalized
mass parameters
the $\bmu$-dependences seen in~\nr{mmH3mu}, \nr{mmU3mu}.



\begin{thebibliography}{99}

\bibitem{rs}
V.A. Rubakov and M.E. Shaposhnikov,
Usp.\ Fiz.\ Nauk 166 (1996) 493 [hep-ph/9603208]. 

\bibitem{rsbb}
K. Rummukainen, IUHET-341 [hep-lat/9608079];
M.E. Shaposhnikov, CERN-TH/96-280 [hep-ph/9610247];
W. Buchm\"uller, DESY 96-216 [hep-ph/9610335];
B. Bergerhoff and C. Wetterich, HD-THEP-96-51 [hep-ph/9611462].

\bibitem{krs}
V.A. Kuzmin, V.A. Rubakov, and M.E. Shaposhnikov,
Phys.\ Lett.\ B 155 (1985) 36;
M.E. Shaposhnikov, Nucl.\ Phys.\ B 287 (1987) 757.

\bibitem{klrs2} 
K. Kajantie, M. Laine, K. Rummukainen and M. Shaposhnikov, 
Nucl.\ Phys.\ B 466 (1996) 189; 
Phys.\ Rev.\ Lett. 77 (1996) 2887.

\bibitem{g}
P.~Ginsparg,
Nucl.\ Phys.\ B 170 (1980) 388;
T. Appelquist and R. Pisarski,
Phys.\ Rev.\ D 23 (1981) 2305;
S. Nadkarni,
Phys.\ Rev.\ D 27 (1983) 917.

\bibitem{fkrs1} 
K. Farakos, K. Kajantie, K. Rummukainen and M. Shaposhnikov, 
Nucl.\ Phys.\ B 425 (1994) 67.

\bibitem{klrs1}
K. Kajantie, M. Laine, K. Rummukainen and 
M. Shaposhnikov, Nucl.\ Phys.\ B 458 (1996) 90. 

\bibitem{bn}
E. Braaten and A. Nieto, 
Phys.\ Rev.\ D 51 (1995) 6990; 53 (1996) 3421.

\bibitem{ck}
J.M. Cline and K. Kainulainen, 
CERN-TH/96-76 [hep-ph/9605235].

\bibitem{lo}
M. Losada, RU-96-25 [hep-ph/9605266]; hep-ph/9612337;
G.R. Farrar and M. Losada, RU-96-26 [hep-ph/9612346].

\bibitem{ml}
M. Laine, Nucl.\ Phys.\ B 481 (1996) 43 [hep-ph/9605283].

%
\bibitem{mgi}
S. Myint, 
Phys.\ Lett.\ B 287 (1992) 325;
G.F. Giudice,  Phys.\ Rev.\ D 45 (1992) 3177. 

\bibitem{eqz} 
J.R. Espinosa, M. Quir\'os and F. Zwirner, 
Phys.\ Lett.\ B 307 (1993) 106.

\bibitem{beqz} 
A. Brignole, J.R. Espinosa, M. Quir\'os and F. Zwirner, 
Phys.\ Lett.\ B 324 (1994) 181. 

\bibitem{cqw}
M. Carena, M. Quir\'os and C.E.M. Wagner,
Phys.\ Lett.\ B 380 (1996) 81.

\bibitem{dggw}
D. Delepine, J.-M. G\'erard, R. Gonzalez Felipe and
J. Weyers, Phys.\ Lett.\ B 386 (1996) 183.

\bibitem{e}
J.R. Espinosa, Nucl.\ Phys.\ B 475 (1996) 273.

\bibitem{ts}
D. Land and E.D. Carlson, 
Phys.\ Lett.\ B 292 (1992) 107;
A. Hammerschmitt, J. Kripfganz and M.G. Schmidt, 
Z.\ Phys.\ C 64 (1994) 105.

\bibitem{leip} 
E.-M. Ilgenfritz, J. Kripfganz, H. Perlt and A. Schiller,
Phys.\ Lett.\ B 356 (1995) 561; 
M. G\"urtler, E.-M. Ilgenfritz, 
J. Kripfganz, H. Perlt and A. Schiller,
hep-lat/9512022; hep-lat/9605042.

\bibitem{desylattice}
F. Csikor, Z. Fodor, J. Hein, A. Jaster and I. Montvay, 
Nucl.\ Phys.\ B 474 (1996) 421.

\bibitem{ccb}
A. Kusenko, P. Langacker and G. Segre, 
Phys.\ Rev.\ D 54 (1996) 5824.

\bibitem{kls1} 
J. Kripfganz, A. Laser and M.G. Schmidt, 
Phys.\ Lett.\ B 351 (1995) 266.

\bibitem{pdp}
K. Farakos, K. Kajantie, K. Rummukainen and 
M. Shaposhnikov, Nucl.\ Phys.\ B 442 (1995) 317; 
W. Buchm\"uller, Z. Fodor and A. Hebecker,
Nucl.\ Phys.\ B 447 (1995) 317.

\bibitem{desy}
D. B\"odeker, W. Buchm\"uller, Z. Fodor, and T. Helbig,
Nucl.\ Phys.\ B 423 (1994) 171.

\bibitem{kls2} 
J. Kripfganz, A. Laser and M.G. Schmidt, 
HD-THEP-95-53 [hep-ph/9512340].

\bibitem{knpr}
F. Karsch, T. Neuhaus, A. Patk\'os and J. Rank, 
Nucl.\ Phys.\ B 474 (1996) 217.

\bibitem{phtw}
O. Philipsen, M. Teper and H. Wittig,
Nucl.\ Phys.\ B 469 (1996) 445.

\bibitem{dkls}
H.-G. Dosch, J. Kripfganz, A. Laser and M.G. Schmidt, 
Phys.\ Lett.\ B 365 (1996) 213.

\bibitem{kk}
K. Kajantie, Phys.\ Lett.\ B 285 (1992) 331; 
J. Ignatius, K. Kajantie, H. Kurki-Suonio and M. Laine,
Phys.\ Rev.\ D 50 (1994) 3738.

\bibitem{eikr}
K. Enqvist, J. Ignatius, K. Kajantie and K. Rummukainen,
Phys.\ Rev.\ D 45 (1992) 3415.

\bibitem{moq}
J.M. Moreno, D.H. Oaknin and M. Quir\'os, 
IEM-FT-127/96 [hep-ph/9605387];
IEM-FT-146/96 [hep-ph/9612212].

\bibitem{jkp}
A.~Jakov\'ac, K.~Kajantie and A.~Patk\'os, 
Phys.\ Rev.\ D 49 (1994) 6810;
A.~Jakov\'ac and A.~Patk\'os, 
Phys.\ Lett.\ B 334 (1994) 391.

\bibitem{hl}
S.-Z. Huang and M. Lissia,
Nucl.\ Phys.\ B 438 (1995) 54.

\bibitem{jw}
R. Jackiw, Phys.\ Rev.\ D 9 (1974) 1686;
R. Fukuda and E. Kyriakopoulos, Nucl.\ Phys.\ B 85 (1975) 354.

\bibitem{ae}
P. Arnold and O. Espinosa, 
Phys.\ Rev.\ D 47 (1993) 3546;
50 (1994) 6662 (E).



\end{thebibliography}
\end{document}